\documentclass[prd,showpacs,showkeys, nofootinbib]{revtex4}
\usepackage{amssymb}

\begin{document}

\title{Voigt transformations in retrospect: missed opportunities?}

\author{Olga~Chashchina}
\affiliation{\'{E}cole Polytechnique, Palaiseau, France}
\email{chashchina.olga@gmail.com}
\author{Natalya~Dudisheva}
\affiliation{Novosibirsk State University, 630 090, Novosibirsk, Russia} 
\email{dudyshevan@mail.ru}
\author{Zurab K.~Silagadze}
\affiliation{Novosibirsk State University  and Budker Institute of Nuclear
Physics, 630 090, Novosibirsk, Russia.}
\email{silagadze@inp.nsk.su}

\begin{abstract}
The teaching of modern physics often uses the history of physics as a
didactic tool. However, as in this process the history of physics
is not something studied but used, there is a danger that the
history itself will be distorted in, as Butterfield calls it,
a ``Whiggish'' way, when the present becomes the measure of the past.

It is not surprising that reading today a paper written more than 
a hundred years ago, we can extract much more of it than was actually thought 
or dreamed by the author himself. We demonstrate this Whiggish approach on 
the example of Woldemar Voigt's 1887 paper. From the modern perspective, 
it may appear that this paper opens a way to both the special relativity and 
to its anisotropic Finslerian generalization which came into the focus only 
recently, in relation with the Cohen and Glashow's very special relativity 
proposal.

With a little imagination, one can connect Voigt's paper to the notorious 
Einstein-Poincar\'{e} priority dispute, which we believe is a Whiggish late 
time artifact. We use the related historical circumstances to give a broader 
view on special relativity, than it is usually anticipated.
\end{abstract}


\maketitle

\section{Introduction}
Sometimes Woldemar Voigt, a German physicist, is considered as ``Relativity's 
forgotten figure'' \cite{I-1}. Although his contribution was not quite 
forgotten and cited ``by those who have taken the trouble to read the original 
literature'' \cite{I-2}, it is a matter of fact that this contribution had no 
influence at all on the development of special relativity. Therefore, in our 
opinion, it is more appropriate in this case to talk about missed 
opportunities in the sense of Dyson \cite{I-3}, than about forgotten 
contribution.

Nearly two decades before the vigorous development of special relativity has
started, in 1887 Woldemar Voigt published an article on the Doppler effect in 
which some fundamental principles underlying the relativity theory were 
anticipated. Name\-ly, he was the first who used Einstein's second postulate 
(universal speed of light) and the restricted form of the first postulate 
(invariance of the wave equation  when changing the inertial reference system) 
to show that the Doppler shift of frequency was incompatible with Newtonian 
absolute time and required a relative time identical with the Lorentz's local 
time introduced later.

The modern reader may wonder why this paper published by Voigt remained 
unnoticed and played no role in the emergence and development of special 
relativity. The simple answer is that we are reading too much in too little: 
what we see in Voigt's paper today is affected by our modern scientific 
knowledge and it doesn't coincide with the readings of Voigt himself and his 
contemporaries. Because it is in general true that a perception of the world 
by human beings, and of scientific papers in particular, depends on everybody's
background shaped by historical circumstances. As put in words by English poet 
William Blake, in his idiosyncratic and provocative form, ``A fool sees not 
the same tree as the wise man sees''. 

Although often ``it is considered legitimate, if not necessary, that
the historian should 'intervene' in the past with the knowledge that he 
possesses by virtue of his placement later in time'' \cite{J-1}, such an 
approach is dangerous and produces what has been called ``Whig'' history 
\cite{J-2,J-3}. In Whig historiography ``what one considers significant in 
history is precisely what leads to something deemed significant today''
\cite{J-3}, so to say, the present becomes the measure of the past.
However, ``the study of the past with one eye, so to speak, upon the present 
is the source of all sins and sophistries in history, starting with the 
simplest of them, the anachronism'' \cite{J-2}. In the present paper our 
objective is not the research in the history of special relativity. Instead
we use the known historical facts and try to properly contextualize the 
original works and thus, from one side, avoid a Whig approach to the history
of special relativity while using it in teaching, and from another one, frame 
and sharpen modern approach to key problems of special relativity, because one 
danger of Whiggish approach is that under its influence students can develop 
an anachronic and archaic perceptions of key features of special relativity.  

The relativity of perception of scientific papers is valid, of course, for
Poincar\'{e}'s far more important contributions too and, in our opinion, is in 
the base of the pointless Einstein-Poincar\'{e} priority dispute. 

This priority dispute originated because history and heritage (in the 
terminology of Grattan-Guinness \cite{J-4,J-5}) were illegitimately 
intermingled by those whose ``normal attention to history is concerned with 
heritage: that is, how did we get here? Old results are modernized in order 
to show their current place; but the historical context is ignored and thereby 
often distorted'' \cite{J-5}.

We argue that in reality there is no priority problem in the history of
special relativity, and this priority dispute originated at later times from
the retrospective reading of Poincar\'{e}'s papers and ignoring the historical 
context. However, we use this artificial priority dispute, inspired by the 
Wiggish approach to the history  of special relativity, to present 
a broad modern perspective on special relativity which we hope will be 
helpful. Besides, these modern developments and views show
that this priority dispute is no longer relevant as far as basic aspects of
special relativity are concerned. 

Moreover, we present a detailed derivations of the anisotropic generalization 
of special relativity by Lalan, Alway and Bogoslovsky and of the corresponding 
Finslerian metric. This generalization is characterized by one extra parameter 
$b$ which has the same conceptual status as the cosmological constant and 
thus, is expected to be very small. Formally, $b\to 0$ limit of the 
anisotropic special relativity is the very special relativity of Cohen and 
Glashow which is an active research topic today. 

The paper is organized as follows. In the next section some circumstances
related to the emergence of special relativity is considered. In particular
we are interested in to find out why the role played by Poincar\'{e} was 
not properly acknowledged at that time by his contemporaries. Our hypothesis
is that this happened because Poincar\'{e}'s approach required a higher 
level of mathematical education than the majority of physicists had at that
time. Minkowski belonged to a few who were in a position to duly appreciate
Poincar\'{e}'s contribution. However he had a psychological reasons to 
downplay the importance of  Poincar\'{e}'s work because otherwise his own
contribution might appear as derivative of Poincar\'{e}'s ideas. We think 
that Minkowski's (wrong) conviction that it was Voigt whom the priority of 
the discovery of the Lorentz transformations should be attributed played 
a dramatic role in his perilous decision not to mention Poincar\'{e} in his
highly influential 1908 Cologne lecture {\it Space and Time}. Thus we devote
the third section to the so-called Voigt transformations, obtained by Voigt 
in 1887, while studying the Doppler effect, well before Lorentz, Einstein and 
Poincar\'{e}'s pioneering works. Our aim in this section is to convey the 
spirit of Voigt's reasoning, not to repeat his derivation. Therefore our 
approach differs in some respect from Voigt's original one and sometimes is
even less general. Then we perform intentionally Whiggish analysis of the
Voigt's transformations and demonstrate that the modern reader can find a lot
of content in it which was impossible for Voigt and his contemporaries. In 
the forth section we provide more modern derivation of Voigt's transformations
which reveals tacit assumptions behind Voigt's original derivation and
later derivations of Lorentz transformations by Poincar\'{e} and Einstein.
Relaxation of some of these tacit assumptions opens a way for interesting 
generalizations of special relativity and in the fifth section we consider one
such generalization, namely the anisotropic special relativity of 
Lalan-Alway-Bogoslovsky, and provide some arguments why we believe the 
dimensionless parameter of this generalization should be very small but 
nevertheless the nonzero value of this parameter is more natural than the very
special relativity recently in the focus of research. In the concluding sixth
section we return to the Einstein-Poincar\'{e} priority dispute and provide
further arguments that this priority dispute is a result of Whiggish approach
to the history of special relativity. We also argue that modern developments
of special relativity makes this priority dispute pointless for another reason
too: many key features of special relativity which by founders of this theory
were considered as crucial and revolutionary doesn't seem so from the modern 
perspective.

\section{Voigt --- a missing link in the Einstein-Poincar\'{e}
mysterious connection?}
It is unfortunate that Einstein didn't cite anybody in his seminal paper
on the foundations of special relativity \cite{1}. On one hand, this 
circumstance can induce a wrong impression that the creation of special
relativity was a miraculous event, a single strike of a young genius, at that 
time a 26-year-old junior clerk at the Swiss patent office in Bern who 
worked it out in his spare time within a few months in complete scientific
isolation. 

Einstein's own account \cite{2} seems to confirm this simplified version
of how the special relativity was created. He recalls that after reading
the Lorentz's 1895 monograph {\it Treatise on a Theory of Electrical and 
Optical Phenomena in Moving Bodies} he spent almost a year in vain trying to 
resolve a difficult problem of how to reconcile the invariance of the velocity 
of light, implied by the validity of Maxwell equations in the reference frame
of the moving body, to the addition rule of velocities used in mechanics.
Then suddenly during a conservation with his friend Michele Besso a solution
came to him: ``An analysis of the concept of time was my solution. Time cannot
be absolutely defined, and there is an inseparable relation between time and
signal velocity. With this new concept, I could resolve all the difficulties 
completely for the first time. Within five weeks the special theory of 
relativity was completed'' \cite{2}.

We have no reason not to believe Einstein. Rindler argues that ``contrary 
claims notwithstanding, apparently no one before Einstein in 1905 ventured the 
idea that all clocks in motion might go slow'' \cite{3}. It seems very natural
that this startling new idea psychologically played a role of a trigger for
Einstein to develop his version of special relativity. This version turned out
very influential and determined the route of special relativity for years 
ahead.

Nonetheless, historically the genesis of the special relativity theory was 
a long process involving at least two more key players: Lorentz and 
Poincar\'{e}. Besides, ``the construction of the special theory of relativity 
did not end with Einstein's papers of 1905. Some features that today's 
physicists judge essential were added only later'' \cite{4}. Many prominent
physicists participated in this process of shaping the special relativity to 
its modern form, most notably Hermann Minkowski, Arnold Sommerfeld, Alfred 
Robb, Max Planck and Max von Laue, to name a few.

Einstein's paper \cite{1} is undoubtedly a strike of a genius. It gives an 
extraordinary clear presentation of fundamentally new concepts about space
and time which shattered the ideas which existed before. However, there is 
a mystery here. As observed by Dyson, ``when the great innovation appears, 
it will almost certainly be in a muddled, incomplete and confusing form. 
To the discoverer himself it will be only half-understood; to everybody else 
it will be a mystery'' \cite{5}. How can we reconcile the extreme clarity
of Einstein's presentation with Dyson's observation?

First of all, Einstein's great discovery has not  appeared out of nowhere.
As Einstein himself wrote in 1955 ``there is no doubt, that the 
\underline{special} theory of relativity, if we regard its development
in retrospect, was ripe for discovery in 1905. Lorentz had already observed 
that for the analysis of Maxwell's equations the transformations which later 
were known by his name are essential, and Poincar\'{e} had even penetrated 
deeper into these connections'' \cite{6}. Although, as the known historical 
documents witness \cite{6,7}, Einstein's version of special relativity is
solely his own long-sought creature, barely influenced by whatever background 
reading he had at that time, undoubtedly Einstein benefited from his scientific
background in this process, at least unconsciously: Einstein received good 
education for his time. Although ``the scientific courses offered to him in 
Z\"{u}rich soon seemed insufficient and inadequate, so that he habitually cut 
his classes. His development as a scientist did not suffer thereby. With 
a veritable mania for reading, day and night, he went through the works
of the great physicists --- Kirchhoff, Hertz, Helmholtz, F\"{o}ppl'' \cite{6}.
Probably we should add Poincar\'{e} to this list as it is known that 
a collection of essays {\it a Science et l'Hypoth\'{e}se} published by 
Poincar\'{e} in 1902 deeply impressed young Einstein and his friends from
Olympia Academy and kept them ``breathless for weeks on end''  \cite{6A}.

As Einstein cites nobody in his paper \cite{1}, we don't know for sure what
contemporary papers, if any, influenced him while writing his masterwork 
\cite{1} and whether his development of special relativity was eased by his 
readings. Letters of Einstein, written between 1898 and 1902, confirm \cite{8} 
his later recollections that he was engaged in the problems of electrodynamics 
of moving bodies many years before writing his epoch-making article \cite{1}. 
However, surviving correspondence sheds very little light on what happened in 
the crucial years between 1902 and 1905. ``Whatever reading and writing he may 
have done at this time, Einstein published nothing on the subject of optics 
and electrodynamics of moving bodies for 3.5 years'' \cite{6,8}.

When Einstein was asked in 1955 whether Poincar\'{e} had had any influence
on his development of special relativity, he answered \cite{6}:
``Concerning myself, I knew only Lorentz's important work of 1895 --- 
'La th\'{e}orie \'{e}lectromagn\'{e}tique de Maxwell' and 'Versuch einer 
Theorie der elektrischen und Optischen Erscheinungenin bewegten K\"{o}rpern' 
--- but not Lorentz's later work, nor the consecutive investigations by 
Poincar\'{e}. In this sense my work of 1905 was independent. The new feature 
of it was the realization of the fact that the bearing of the Lorentz 
transformation transcended its connection with Maxwell's equations and was 
concerned with the nature of space and time in general. A further new result 
was that the 'Lorentz invariance' is a general condition for any theory''.

Einstein is not very accurate here. The universal character of Lorentz 
invariance, the fact that it is a general requirement for all laws of 
physics, not merely a property of  Maxwell's equations, was, no doubt, 
anticipated by Poincar\'{e} \cite{4,9}. Moreover, there is still another 
mystery. From the modern perspective, it may appear that the Einstein's 
synchronization procedure via light signals and not absolute nature of 
simultaneity were also Poincar\'{e}'s inventions. Already in 1898,
``Poincar\'{e} had presented exactly the same light signaling and clock 
synchronization thought experiment that would later be found in Einstein's 
1905 relativity paper'' \cite{10}, although Poincar\'{e}'s presentation is 
without any mention of the relativity principle and Lorentz's local time.

Two years later in his lecture ``Lorentz's theory and the principle of 
reaction'' Poincar\'{e} used his light signaling and clock  synchronization
thought experiment to explain the physical meaning of the Lorentz's local 
time \cite{10}. It is true however that Poincar\'{e} never abandoned the
concept of {\ae}ther and considered the above mentioned relativity of 
simultaneity as only an apparent, not genuine, physical effect related to
the fact that we have synchronized the clocks in the moving frame as if the 
light velocity were the same in all directions, like what happens in the 
preferred {\ae}ther frame. For Poincar\'{e}, only clocks so synchronized in the
{\ae}ther frame show the real time, despite the fact that he was fully aware 
that, because of the relativity principle, it was impossible to experimentally
find out that the local time defined in the moving frame was not the real time.
In 1902 letter to the Nobel committee to nominate Lorentz for the Nobel prize
in Physics, which he indeed was awarded, Poincar\'{e} praises very highly
Lorentz's ``most ingenious invention'' of ``local time'' and writes: 
``Two phenomena happening in two different places can appear simultaneous even 
though they are not: everything happens as if the clock in one of these places 
were late with respect to that of the other, and as if no conceivable 
experiment could show evidence of this discordance'' \cite{10}.

Could Poincar\'{e}'s insights about simultaneity and clock synchronization
somehow influenced Einstein? This is not an easy question.
We know from Einstein himself that he had read Lorentz's 1895 treatise and
``was therefore aware of the local time and the role it served in preserving 
the form of the Maxwell-Lorentz equations to first order'' \cite{4}. We also
know that he and his friends from Olympia Academy were fascinated by
Poincar\'{e}'s {\it La science et l'hypoth\'{e}se}, which contained the 
relativity principle and a brief critical discussion of simultaneity. ``He may
also have known the exact form of the Lorentz transformations, for in 1904 
several German theorists commented on them in journals that he regularly read''
\cite{4}. For Torretti this is enough to conclude that ``there is no doubt
that Einstein could have drawn inspiration and support for his first work on
relativity from the writings of Poincar\'{e}'' \cite{11}. However matters are
not as simple, because Einstein denies such an influence and we have no reason
not to believe him. Knowing the paradoxical nature of human memory and 
psychology, we cannot exclude that the true story was hidden for Einstein too.

``It is certain that sometime before 17 May 1906, Einstein did read 
Poin\-car\'{e}'s 1900 paper --- on that day Einstein submitted a paper of his 
own that explicitly used the contents of Poincar\'{e}'s article (though not 
local time)'' \cite{12}. This was the first and last occasion Einstein ever
cited Poincar\'{e} in the context of relativity (in connection with the 
$E=mc^2$ formula).

Similar ideas about clock synchronization and physical definition of the local
time can be also found (in German) in Emil Cohn's November 1904 article 
``Toward the electrodynamics of moving systems'' \cite{12}. Cohn's views are
even closer to Einstein's than to  Poincar\'{e}'s, as he rejected the 
{\ae}ther, preferring ``the vacuum''. We know that sometime before 
25 September 1907 Einstein did read the Cohn's paper --- at that day he sent 
to the editor of a journal his review paper on relativity which contains 
a footnote-compliment: ``The pertinent studies of E.~Cohn also enter into 
consideration, but I did not make use of them here'' \cite{12}. However, again 
we have no evidence that Einstein knew the Cohn's work prior his founding 1905 
paper. How can we then explain a strange fact that ``Einstein's reasoning in 
1905 regarding synchronization of clocks using light signals is incredibly 
similar to Poincar\'{e}'s'' \cite{13}? Probably Chavan or Solovine (members of 
the Olympia Academy), who were fluent in French, reported about Poincar\'{e}'s
1898 paper (in French) ``the measurement of time'' at an Olympia Academy
session \cite{12}. Even more interesting guess about what could have happened 
then is suggested in \cite{10}.

Norton argues that it is quite possible that thoughts of clocks 
synchronization by light signals played no essential role in Einstein's 
discovery of the relativity of simultaneity: ``A plausible scenario is that 
Einstein was compelled to the Lorentz transformation for space and time as 
a formal result, but needed some way to make its use of local time physically
comprehensible. Thoughts of light signals and clock synchronization would then 
briefly play their role. It is also entirely possible that these thoughts 
entered only after Einstein had become convinced of the relativity of 
simultaneity; that is, they were introduced as an effective means of conveying 
the result to readers of his 1905 paper and convincing them of it. In both 
cases, thoughts about the light signals and clock synchronization most likely 
played a role only at one brief moment, some five to six weeks prior to the 
completion of the paper, at the time that Einstein brought his struggle with 
him to a celebrated meeting with his friend Michele Besso'' \cite{14}. Indeed, 
five to six weeks before completing his relativity paper, Einstein wrote to 
his friend Conrad Habicht: ``The fourth paper is only a rough draft at this 
point, and is an electrodynamics of moving bodies which employs a modification 
of the theory of space and time'' \cite{10}. It seems that at that time 
Einstein already discovered the relativity of simultaneity, as he speaks of 
modification of the theory of space and time, but yet had not found a good way 
how to present his new discoveries. Then there was a ``celebrated meeting'' 
with his friend Michele Besso. At the end of his relativity paper, Einstein 
writes ``In conclusion I wish to say that in working at the problem here dealt 
with I have had the loyal assistance of my friend and colleague M. Besso, and 
that I am indebted to him for several valuable suggestions'' \cite{1}. What 
was Besso's role in completing Einstein's relativity paper? Granek suggests 
``that in the Einstein-Besso 'celebrated meeting' Besso indirectly and 
unnoticeably conveyed Poincar\'{e}'s light signaling and clock synchronization 
thought experiment, without necessarily knowing that this thought experiment 
originated in Poincar\'{e}'s works'' \cite{10}. Probably Besso was told about 
this light signaling and clock synchronization thought experiment by another 
member of Olympia Academy who read Poincar\'{e}'s papers (as was already 
mentioned, it is possible that Solovine read Poincar\'{e}'s 1898 paper). Then 
he conveyed to Einstein a loose idea about the light signaling and clock 
synchronization intermingled with other ideas \cite{10}. No wonder that 
Einstein afterwords refused any connection with Poincar\'{e}, because he was 
convinced that he came to the idea by himself, with the help of his friend 
Besso, and had not realized that in fact this was Poincar\'{e}'s idea which 
came to him in such a bizarre way. 

We think this is the best we can do on the question whether Poincar\'{e}'s
or Cohn's papers influenced Einstein's reasoning. ``The limits of 
reconstruction are evident, and would be even if it were written in stone that
Einstein had seen one of these papers'' \cite{12}.

However, the real enigma is not the absence of Poincar\'{e}'s name in 
Einstein's 1905 paper. As we have seen above, we can imagine a plausible
explanation of this fact. The real enigma is why Poincar\'{e}'s contribution
to relativity was downplayed by his contemporaries. 

From the modern perspective, Poincar\'{e}'s treatment of the Lorentz group in
his last pre-Einstein papers \cite{15,16} is extraordinary advanced and modern
\cite{9,17}. In some sense, Poincar\'{e} was far ahead of his time. Perhaps
very few physicists (if any) at that time had enough mathematical background
to duly appreciate Poincar\'{e}'s achievements. Besides, Poincar\'{e}'s 1906
Rendiconti paper \cite{16} was both ``rather long (47 pages) and technically 
quite complex, especially in the final section'' where ``many of the key new 
results of Poincar\'{e} on Relativity are contained'' \cite{18}. However,
Hermann Minkowski was undoubtedly among those who were capable to appreciate
Poincar\'{e}'s writings. And a blatant fact, which leaves us almost speechless,
is that Poincar\'{e}'s name was never mentioned in Minkowski's famous Cologne
lecture ``Raum und Zeit''.

Minkowski's September 21st, 1908, lecture ``Space and Time'' was a crucial 
event in the history of relativity \footnote{Annalen der Physik in 2008
(vol. 17, issues 9 and 10, pp. 613-852) published a collection of articles 
written to celebrate  the 100th anniversary of the Minkowski's 1908 lecture.
We cite two articles from this  collection (by Damour and by Kastrup) in this
manuscript. However, readers can find other articles from this  collection 
also interesting.}. In a sense, the appearance of the 
Einstein's paper \cite{1} was an unexpected and peculiar event which changed 
the natural flow of ideas of contemporary physics. 
Suppose Einstein had not existed. How would our understanding of space-time 
physics have developed then? In an interesting essay \cite{19} 
L\'{e}vy-Leblond tried to answer this question. His conclusion is that ``apart 
from a few details perhaps, the formalism, that is, the notations and 
equations, would be very similar to ours today. But the language and, 
underneath, the words and the ideas we would use could  be rather different. 
Many familiar terms, beginning with 'relativity' itself, might be absent from 
our vocabulary'' \cite{19}. In this respect, Minkowski's Cologne lecture 
constitutes the shift of focus from reference frames and relativity of lengths 
and time intervals towards the modern concepts of symmetry and associated 
space-time geometry. After the Cologne lecture, relativity was gradually but 
firmly embarked  on the royal way leading to the modern understanding of 
special relativity as the Minkowskian chronogeometry of the 4-dimensional 
flat space-time.

Initially Einstein was skeptical about Minkowski's innovations and called
them ``superfluous learnedness'' \cite{20}. According to Sommerfeld's 
recollections, he said  on one occasion: ``Since the mathematicians have 
invaded the relativity theory, I do not understand it myself any more''
\cite{21}. However, as early as 1912 Einstein fully appreciated the strength
of Minkowski's 4-dimensional approach. At that year he wrote to  
Sommerfeld: ``I am now occupied exclusively with the gravitational problem, 
and believe that I can overcome all difficulties with the help of a local 
mathematician friend [Marcel Grossmann]. But one thing is certain, never before
in my life have I troubled myself over anything so much, and that I have 
gained great respect for mathematics, whose more subtle parts I considered 
until now, in my ignorance, as pure luxury! Compared with this problem, the 
original theory of relativity is childish'' \cite{21}.

The influence of the Cologne lecture was enormous. Its published version 
``sparked an explosion of publications in relativity theory, with the number 
of papers on relativity tripling between 1908 (32 papers) and 1910 
(95 papers)'' \cite{22}. The response to the Minkowski's lecture was 
overwhelmingly positive on the part of mathematicians, and more mixed 
on the part of physicists --- only in the 1950s their attitude began to 
converge toward Minkowski's space-time view \cite{22}.

Due to the importance of the Cologne lecture and to the impact it had on the 
historical development of special relativity, a sad omission of Poincar\'{e}'s 
name from it played, in our opinion, a crucial role in the subsequent 
downplaying of Poincar\'{e}'s contribution to relativity by contemporary 
physicists for a long time.

Some leading scientists remarked the absence of Poincar\'{e}'s name in the
Cologne lecture and even tried somehow to correct injustice. In particular,
``in the notes he added to a 1913 reprint of this lecture, Sommerfeld 
attempted to right the wrong by making it clear that a Lorentz-covariant law 
of gravitation and the idea of a 4-vector had both been proposed earlier by 
Poincar\'{e}'' \cite{23}. Though Sommerfeld acknowledges Poincar\'{e}'s 
contribution only partially, the young Pauli does ``a better job'' \cite{18} 
in giving Poincar\'{e} a due credit in his famous review article on the theory 
of relativity \cite{24}. In fact, this happened under Felix Klein's
direct influence. In a letter of acknowledgment of the receipt of the 
manuscript Klein advises Pauli to pay great attention to historical facts
and in particular emphasizes Poincar\'{e}'s priority before Einstein in
recognizing the group properties of the Lorentz transformations \cite{25}.
However, these efforts were insufficient. As the citation record shows 
\cite{26}, Poincar\'{e}'s papers \cite{15,16} were rarely cited during the
founding period of relativity 1905-1912 and thus had virtually no effect
on the development of special relativity (in this time period Einstein's
paper \cite{1} was cited about five times more frequently than Poincar\'{e}'s 
papers \cite{15,16}).

Why did Minkowski not mention  Poincar\'{e}? Minkowski was certainly aware of
Poincar\'{e}'s papers. One year before the celebrated lecture in Cologne,
in his lecture to  the G\"{o}ttinger Mathematischen Gesellschaft, the text of 
which was published posthumously in 1915 through the efforts of Sommerfeld, 
Minkowski frequently and positively cites  Poincar\'{e}. In fact Poincar\'{e}
was the third most cited author (cited six times), after Planck (cited eleven
times) and Lorentz (cited ten times), in this lecture \cite{18}. For 
comparison, Einstein was cited only twice. How can we then explain the 
complete disregard of Poincar\'{e}'s contribution only one year later? In 
absence of any historical document resolving this conundrum, only speculations 
on Minkowski's motivations may be entertained, as was done in \cite{18} and
\cite{23}. Maybe the following circumstances played a psychological role.

According to Max Born's recollections, later after the Cologne lecture 
Minkowski told him that ``it came to him as a great shock when Einstein 
published his paper in which the equivalence of the different local times of 
observers moving relative to each other was pronounced; for he had reached the 
same conclusions independently but did not publish them because he wished 
first to work out the mathematical structure in all its splendor'' \cite{21}.
We may suppose that the discovery of the 4-dimensional formalism was also
a result of this process of working out the mathematical structure behind the
Lorentz transformations and was made by Minkowski independently of 
Poincar\'{e}'s 1905 papers. ``If that reconstruction is correct, he must have 
been all the more eager, when he later realized that he had been preceded by 
Poincar\'{e}, to find reasons for downplaying Poincar\'{e}'s work'' \cite{18}.
So, perhaps, in 1907 Minkowski had only a background reading of Poincar\'{e}'s
papers. Then, as he prepared for the Cologne lecture, he more thoroughly
studied Poincar\'{e}'s papers and realized that if he ``had chosen to include 
some mention of Poincar\'{e}'s work, his own contribution may have appeared 
derivative'' \cite{23}. However, in our opinion, to make the decision to 
exclude Poincar\'{e}'s name from the Cologne lecture Minkowski needed some 
serious reason to psychologically justify such an unfair omission. Maybe 
another oddity of the Cologne lecture gives a clue what was this reason. 

In the Cologne lecture Minkowski never mentions Lorentz transformations (so 
named by Poincar\'{e} in \cite{15,16}), instead he refers to transformations 
of the group $G_c$. Even more surprisingly, on December 21, 1907, 
Minkowski talked to the G\"{o}ttingen scientific society and subsequently in 
April 1908 published its contents in {\it G\"{o}ttinger Nachrichten} as a 
technical paper {\it Die Grundgleichungen f\"{u}r die electromagnetischen 
Vorg\"{a}nge in bewegten K\"{o}rpern} where one can find all the results of 
the Cologne lecture presented with great details \cite{18,21}. In this paper 
Minkowski quotes Poincar\'{e} only twice and ``in a rather derogatory 
manner'' \cite{18}. In particular,  at the beginning, he mentions Poincar\'{e} 
only as an originator of the name ``Lorentz transformations''. In fact, 
probably, this talk of Minkowski was a beginning of his downplaying of the
Poincar\'{e}'s contribution, as witnessed by the following facts \cite{26A}.  
In it Minkowski gives Lorentz the credit of having found the ``purely 
mathematical fact'' of the Lorentz covariance of the Maxwell equations and 
of having ``created the relativity-postulate''. None of these claims is true 
--- it is Poincar\'{e} who should be credited for these discoveries as 
was explicitly stated by Lorentz himself. The priority in clarifying the link 
between the Lorentz transformations and a new concept of time is fully given 
to Einstein without mentioning Poincar\'{e}'s important insights in this 
question. Overall impression that the reader could infer from this 
historically important publication might be that Minkowski  is ``suggesting 
that the main (if not the only) contribution by Poincar\'{e} is to have given 
the Lorentz transformations \ldots their name'' \cite{26A}.

The reason for suppression of the name ``Lorentz transformations'' in the 
Cologne lecture is unknown, but very probably \cite{23} it was linked to 
Minkow\-ski's discovery that the essential application of the Lorentz symmetry 
in optics goes back to Woldemar Voigt's 1887 paper \cite{27}.

Thanks to a letter sent by Minkowski to David Hilbert in 1889, we know that
Minkowski was studying Voigt's treatment of elasticity \cite{28A}. However,
it is not known whether Minkowski read Voigt's 1887 paper on the Doppler 
principle at that time. Probably he did. At least such a suggestion seems 
tenable \cite{28}. Even if he read the paper, it seems he had not recognized 
the potential of Voigt's transformations. In his letter of 1889 to Hilbert
he mentions that ``abstract speculation was not sufficient for physical 
understanding'' and that ``it was inconceivable to him that anyone would 
develop mathematical equations only with the hope that someone might later 
demonstrate their utility'' \cite{28A}. 

However, it is quite possible and even more plausible that Minkowski became 
aware of the Voigt's 1887 Doppler principle paper only later and his above 
given quotations concern to another paper by Voigt {\it Theoretische 
Studien \"{u}ber die Elasticit\"{a}tsverh\"{a}ltnisse der Krystalle} published
also in 1887 \cite{28B}. Interestingly, Walter gives a slightly different
translation of Minkowski's quotation: ``how anyone can apply odd calculations 
in the hope that later, perhaps, someone will be found who can get something 
out of it'' \cite{28B} \footnote{The German original can be found in 
\cite{28BB} and goes as follows: ``Es ist mir ganz unbegreiflich, wie jemand 
w\"{u}ste Rechnungen ansetzen kann, in der Hoffnung da{\ss} sich sp\"{a}ter 
vielleicht jemand findet, der Nutzen daraus zu ziehen im Stande ist''.}.

Although Voigt and Lorentz were in correspondence since 1883,  Lorentz was
unaware of Voigt's Doppler principle paper and it was not until 1908 that
Voigt sent him a reprint of this paper. In a letter of acknowledgment of the 
receipt of the preprint from July, 1908, Lorentz writes ``I would like to 
thank you very much for sending me your paper on Doppler's principle
together with your enclosed remarks. I really regret that your paper has 
escaped my notice. I can only explain it by the fact, that many lectures kept 
me back from reading everything, while I was already glad to be able to work 
a little bit. Of course I will not miss the first opportunity to mention, that 
the concerned transformation and the introduction of a local time has been 
your idea'' \cite{28}.

Voigt and Minkowski were friends at G\"{o}ttingen. So it is possible that
Voigt informed Minkowski about his correspondence with Lorentz in around July, 
1908 and probably that's how Minkowski became aware of the Voigt's 1887 
Doppler principle paper \cite{28}. 

In confirmation of this suggestion Minkowski cites (incorrectly) Voigt in the 
Colog\-ne lecture as the discoverer of the Lorentz symmetry. Namely, he says
``Now the impulse and true motivation for assuming the group $G_{c}$ came from 
the fact that the differential equation for the propagation of light waves in 
empty space possesses the group $G_{c}$'' and adds in the footnote that 
an essential application of this fact can already be found in Voigt's 1887 
paper \cite{28C}. 

Two days after the Cologne lecture, in a discussion that followed the  
Alfred Bucherer's lecture {\it Measurements of Becquerel rays. The Experimental
Confirmation of the Lorentz-Einstein Theory} \footnote{Bucherer's experimental 
results contributed significantly to the physicists adjustment to the theory
of relativity as they refuted earlier results by Walter Kaufmann chalennging
the empirical adequacy of the Lorentz-Einstein theory. ``In the discussion of 
this lecture, Minkowski expressed joy in seeing the 'monstrous' rigid electron 
hypothesis experimentally defeated in favor of the deformable electron of 
Lorentz's theory'' \cite{26}.}, Minkowski again tried to promote
Voigt as the discoverer of Lorentz transformations. He said: ``Historically, I 
want to add that the transformation, which play the main role in the relativity
principle, were first mathematically discussed by Voigt in the year 1887. 
Already then, Voigt drew some consequences with their aid, in respect to the 
principle of Doppler'' \cite{28D}.

Convincing himself that it was Voigt who should be credited for apprehension
of the central role of the Lorentz symmetry, perhaps, it was psychologically 
more easy for Minkowski to decide to omit Poincar\'{e}'s name from the Cologne 
lecture. Very likely, this pernicious decision was further eased by 
Poincar\'{e}'s style of writing who ``habitually wrote in a self-effacing 
manner. He named many of his discoveries after other people, and expounded many
important and original ideas in writings that were ostensibly just reviewing 
the works of others, with 'minor amplifications and corrections'. 
Poincar\'{e}'s style of writing, especially on topics in physics, always gave 
the impression that he was just reviewing someone else's work'' \cite{28E}.
   
Curiously, Voigt's name resurfaces, and in a rather mysterious way, in 
relation with Einstein's paper \cite{1} too (see \cite{29}). In \cite{1} 
Einstein doesn't use Lorentz's and Poincar\'{e}'s nomenclature $(x, y, z, t)
\leftrightarrow (x^\prime, y^\prime, z^\prime, t^\prime)$ while writing the 
final form of Lorentz transformations, but rather the one used in Voigt's 
paper \cite{27}: $(x, y, z, t)\leftrightarrow (\xi, \eta, \zeta, \tau)$. 
We don't know the reason of this weird coincidence. Did Einstein read Voigt's 
paper? Or was it reported on the Olympia Academy session? Probably
not and all this is just a coincidence. After all 
$(\xi_1, \eta_1, \zeta_1)$, $(\xi_2, \eta_2, \zeta_2)$, etc. were used by 
Lorentz to denote locations of the molecules in his 1895 treatise which 
Einstein was well aware of. Poincar\'{e} also uses $(\xi, \eta, \zeta )$, 
though to denote velocity components, in \cite{15,16}.
If you still consider such a coincidence as improbable, like the author of 
\cite{29}, one can remember another amazing coincidence of this kind. The 
following two identities, expressing parametric solutions for representing 
a forth power as a sum of five  forth powers, can be found in Ramanujan's 
third notebook \cite{29A}:
\begin{equation}
(8s^2+40st-24t^2)^4+(6s^2-44st-18t^2)^4+(14s^2-4st-42t^2)^4+ 
(9s^2+27t^2)^4+(4s^2+12t^2)^4=(15s^2+45t^2)^4, \nonumber
\end{equation} 
and
\begin{equation}
(4m^2-12n^2)^4+(3m^2+9n^2)^4+(2m^2-12mn-6n^2)^4+
(4m^2+12n^2)^4+(2m^2+12mn-6n^2)^4=(5m^2+15n^2)^4. \nonumber
\end{equation}  

What is amazing about this identities is that the first identity was published
by Haldeman in 1904 and Ramanujan uses the same notations and precisely the 
same order of the terms as Haldeman. ``One might conclude that Ramanujan saw 
Haldeman's paper or a secondary source quoting it. However, this seems highly 
unlikely as Ramanujan had access to very few journals in India and, 
moreover, Haldeman's paper was published in a very obscure journal. It is also 
quite possible that Ramanujan made his discovery before Haldeman did. Thus, in 
conclusion, the identical notation must be an amazing coincidence'' 
\cite{29A}.

The second identity is also contained in that Haldeman's paper.
However, in this case Ramanujan uses not the same notation as Haldeman.
His notation (but not order of the terms) coincides to what was used by
A.~Martin, another mathematician who published his version of the proof of 
this identity in the same volume in which Haldeman's paper appeared.
``Again, this must be an astonishing coincidence'' \cite{29A}. Probably, 
the same can be said about Einstein-Voigt notation coincidence.

\section{Voigt transformations}
Now let us take a closer look at the Voigt's 1887 Doppler principle paper 
\cite{28,29,30,31,32}. The first impression is that this paper \cite{27} looks 
more like a technical note with messy calculations than an ordinary 
research paper: there is no abstract present nor any explanation of 
the idea and purpose of the paper, and, like Einstein's \cite{1}, it does not 
contain references \cite{31}. Voigt's objective was to derive the Doppler 
effect formula from the invariance of the wave equation
\begin{equation}
\Box\, \Phi(x,y,z,t)=\left(\frac{1}{c^2}\frac{\partial^2}{\partial t^2}-
\Delta \right) \Phi(x,y,z,t)=0,
\label{eq1}
\end{equation} 
under the change of variables
\begin{equation}
\begin{array}{l}
\xi=m_1(V)\,x+n_1(V)\,y+p_1(V)\,z-\alpha_0(V)\, t, \\
\eta=m_2(V)\,x+n_2(V)\,y+p_2(V)\,z-\beta_0(V)\, t, \\
\zeta=m_3(V)\,x+n_3(V)\,y+p_3(V)\,z-\gamma_0(V)\, t, \\
\tau=t-\left [a_0(V)\,x+b_0(V)\,y+c_0(V)\,z\right ],
\end{array}
\label{eq2}
\end{equation} 
which corresponds to the transition from the {\ae}ther rest frame to an 
other frame moving with a constant velocity $\vec{V}$. Voigt doesn't explain 
why such invariance (which implies that the light velocity is the same in 
both frames) should take place, he simply says that the wave equation in the 
new frame has the form
\begin{equation}
\left(\frac{1}{c^2}\frac{\partial^2}{\partial \tau^2}-
\frac{\partial^2}{\partial \xi^2}-\frac{\partial^2}{\partial \eta^2}-
\frac{\partial^2}{\partial \zeta^2}\right) \Phi(\xi,\eta,\zeta,\tau)=0,
\label{eq3}
\end{equation} 
``as it must be'' \cite{31}. Neither he provides any rationale behind the
coordinate transformation (\ref{eq2}).
 
Comparing (\ref{eq1}) and (\ref{eq3}), Voigt obtains a system of nine 
equations for fifteen unknown constants 
$$m_1,\,n_1,\,p_1,\,m_2,\,n_2,\,p_2,\,m_3,\,n_3,\,p_3,a_0,\, b_0,\, c_0,\,
\alpha_0,\,\beta_0,\,\gamma_0.$$
Therefore, he needs some assumptions to solve this troublesome system.
First of all he notices that in the moving frame $S^\prime$ the origin
($x=0,y=0,z=0$) of the stationary frame $S$ moves with the velocity 
$-(\alpha_0,\beta_0,\gamma_0)$. Therefore, $(\alpha_0,\beta_0,\gamma_0)$ can 
be identified with $\vec{V}$ and assuming the standard configuration 
($S^\prime$ moves along the $x$-axis with respect to $S$), we get 
$\alpha_0=V,\,\beta_0=0,\,\gamma_0=0$. The following can then be calculated 
from (\ref{eq2}):
\begin{eqnarray} && 
\frac{\partial}{\partial x}=m_1\,\frac{\partial}{\partial \xi}+
m_2\,\frac{\partial}{\partial \eta}+m_3\,\frac{\partial}{\partial \zeta}-
a_0\,\frac{\partial}{\partial \tau}, \nonumber \\ && 
\frac{\partial}{\partial y}=n_1\,\frac{\partial}{\partial \xi}+
n_2\,\frac{\partial}{\partial \eta}+n_3\,\frac{\partial}{\partial \zeta}-
b_0\,\frac{\partial}{\partial \tau}, \nonumber \\ && 
\frac{\partial}{\partial z}=\;p_1\,\frac{\partial}{\partial \xi}+
\;p_2\;\frac{\partial}{\partial \eta}+\;p_3\;\frac{\partial}{\partial \zeta}-
\;c_0\;\frac{\partial}{\partial \tau}, \;\;\;\;
\frac{\partial}{\partial t}=\frac{\partial}{\partial \tau}-V\,\frac{\partial}
{\partial \xi}.
\label{eq4}
\end{eqnarray}
The invariance of the wave equation implies that $\Box^\prime=\gamma^2(V)\,
\Box$, or
\begin{equation}
\frac{1}{c^2}\frac{\partial^2}{\partial \tau^2}-
\frac{\partial^2}{\partial \xi^2}-\frac{\partial^2}{\partial \eta^2}-
\frac{\partial^2}{\partial \zeta^2}=\gamma^2(V)\left (
\frac{1}{c^2}\frac{\partial^2}{\partial t^2}-\frac{\partial^2}
{\partial x^2}-\frac{\partial^2}{\partial y^2}-\frac{\partial^2}{\partial z^2}
\right ),
\label{eq5}
\end{equation}
where $\gamma^2(V)$ is some constant. In combination with (\ref{eq4}) this 
condition gives ten equations which we organize into three groups (starting 
from this point we somewhat modify Voigt's original derivation in order to 
make it more transparent). Equations of the first group correspond to the 
diagonal elements of the d'Alembert operator and they have the form
\begin{eqnarray} &&
\gamma^2\left [1-(a_0^2+b_0^2+c_0^2)c^2\right ]=1,\;\;\;\;
\gamma^2\left [(m_1^2+n_1^2+p_1^2)-\frac{V^2}{c^2}\right ]=1, \nonumber \\ &&
\gamma^2\left [m_2^2+n_2^2+p_2^2\right ]=1,\hspace*{20mm}
\gamma^2\left [m_3^2+n_3^2+p_3^2\right ]=1.
\label{eq6}
\end{eqnarray}
The second group corresponds to the equations that ensure the vanishing of the 
mixed derivatives $\frac{\partial^2}{\partial \tau \partial \xi}$,
$\frac{\partial^2}{\partial \tau \partial \eta}$ and 
$\frac{\partial^2}{\partial \tau \partial \zeta}$. These equations are
\begin{eqnarray} && 
m_1\,a_0+n_1\,b_0+p_1\,c_0-\frac{V}{c^2}=0,\nonumber \\ &&
m_2\,a_0+n_2\,b_0+p_2\,c_0=0,\;\;m_3\,a_0+n_3\,b_0+p_3\,c_0=0.
\label{eq7}
\end{eqnarray}
The third group of equations ensures the vanishing of the mixed 
derivatives $\frac{\partial^2}{\partial \xi \partial \eta}$,
$\frac{\partial^2}{\partial \xi \partial \zeta}$ and 
$\frac{\partial^2}{\partial \eta \partial \zeta}$. They have the form
\begin{eqnarray} &&
m_1\,m_2+n_1\,n_2+p_1\,p_2=0,\nonumber\\ &&
m_1\,m_3+n_1\,n_3+p_1\,p_3=0,\;\;
m_2\,m_3+n_2\,n_3+p_2\,p_3=0.
\label{eq8}
\end{eqnarray}
As we see, we have thirteen unknowns (including $\gamma^2$) and ten equations.
So, Voigt concludes that three of them can be chosen arbitrarily. The natural
choice is $m_1=1,\,n_1=0,\,p_1=0$, which makes the transformation law for 
$\xi$ identical to the Galilean transformation. Then equations (\ref{eq8})
imply $m_2=m_3=0$ and
\begin{equation}
n_2\,n_3=-p_2\,p_3,
\label{eq9}
\end{equation}
while from equations (\ref{eq6}) we get
\begin{equation}
\gamma^{-2}=1-\frac{V^2}{c^2},
\label{eq10}
\end{equation}
and
\begin{equation}
n_2^2+p_2^2=n_3^2+p_3^2=\gamma^{-2}.
\label{eq11}
\end{equation}
At last, equations (\ref{eq7}) will give us in this case
\begin{equation}
a_0=\frac{V}{c^2},
\label{eq12}
\end{equation}
and
\begin{equation}
n_2\,b_0=-p_2\,c_0,\;\;\;n_3\,b_0=-p_3\,c_0.
\label{eq13}
\end{equation}
Equations (\ref{eq9}) and (\ref{eq13}) can be combined to produce the 
following relations
\begin{equation}
(n_2^2+p_2^2)p_3\,c_0=0,\;\;\;(n_3^2+p_3^2)n_2\,b_0=0,
\label{eq14}
\end{equation}
which in light of (\ref{eq11}) imply
\begin{equation}
p_3\,c_0=0,\;\;\;n_2\,b_0=0.
\label{eq15}
\end{equation}
However, in the limit $V=0$ we should have $n_2(0)=1$ and $p_3(0)=1$.
Therefore, $n_2$ and $p_3$ can't be identically zero and the only way to 
satisfy (\ref{eq15}) is to take $b_0=c_0=0$.

If we introduce two complex numbers $z_1=n_2+ip_2$ and $z_2=p_3+in_3$, the 
equations (\ref{eq9}) and (\ref{eq11}) indicate that $|z_1|=|z_2|=\gamma^{-1}$ 
and that $z_1z_2$ is a real number. Therefore, $z_2=\pm z_1^*$. Only the plus 
sign is acceptable, because $n_2(0)=p_3(0)$. Thus, the most general solution
for $z_1$ and $z_2$ has the form
\begin{eqnarray} &&
n_2=\gamma^{-1}\cos{\phi(V)},\;\;\;\;\;p_2=\gamma^{-1}\sin{\phi(V)},
\nonumber \\ &&
n_3=-\gamma^{-1}\sin{\phi(V)},\;\;\;p_3=\gamma^{-1}\cos{\phi(V)},
\label{eq16}
\end{eqnarray}
where $\phi(V)$ is any function of the velocity $V$ such that $\phi(0)=0$.
The presence of an angle $\phi(V)$ reflects the obvious fact that 
a transformation from the {\ae}ther frame to the moving frame can always be 
superimposed by a rotation around the axis parallel to $\vec{V}$ by an angle
$\phi(V)$. Excluding this trivial possibility of an additional rotation,
we see that the invariance of the wave equation under Voigt's assumptions 
essentially uniquely leads to Voigt transformations:
\begin{equation}
\begin{array}{l}
\xi=x-V\, t, \\
\eta=\gamma^{-1}\,y, \\
\zeta=\gamma^{-1}\,z, \\
\tau=t-\frac{V}{c^2}\,x.
\end{array}
\label{eq17}
\end{equation} 
The inverse transformations have the form
\begin{equation}
\begin{array}{l}
x=\gamma^2\left(\xi+V\, \tau\right), \\
y=\gamma\,\eta, \\
z=\gamma\,\zeta, \\
t=\gamma^2\left(\tau+\frac{V}{c^2}\,\xi\right).
\end{array}
\label{eq18}
\end{equation}  
For a modern reader it is obvious that Voigt was very close to discovering 
special relativity. His transformations reveal the crucial fact that the
invariance of the wave equation is inconsistent with Newtonian absolute time.
In fact, Voigt was the first who discovered ``that a 'natural' clock would
alter its rate on motion'' \cite{33}. More precisely, to be fair, Voigt missed 
this opportunity because he apparently did not realized a great conceptual 
novelty of his transformations. His main objective was the Doppler effect. 
Using his transformations he concludes that the vibration period experiences 
time dilation with the (wrong) factor $\gamma^2$. However, the conclusion that 
all moving clocks would actually behave like this, was not ventured and to 
a great extent ``his paper appears to be a mathematical exercise'' without 
giving any conceptual reason for developing this particular set of 
transformations \cite{33}. 

Voigt transformations (\ref{eq17}) differ from Lorentz transformations only 
by a scale factor. Many physicists, including Minkowski and Lorentz, considered
this difference insignificant and credited Voigt with the discovery of Lorentz 
transformations \cite{31}. For example, Lorentz wrote in 1914: ``These 
considerations published by myself in 1904, have stimulated Poincar\'{e} to
write his article on the dynamics of electron where he has given my name to
the just mentioned transformation. I have to note as regards this that 
a similar transformation has been already given in an article by Voigt 
published in 1887 and I have not taken all possible benefit from it'' 
\cite{9}. It seems Voigt himself finally shared this opinion. Voigt's 1887 
paper was reprinted in {\it Physikalische Zeitschrift} on occasion of the 
tenth anniversary of the principle of relativity in 1915. In the reprinted 
version Voigt included additional comments and in particular when referring to 
his transformations he wrote: ``This is, except for the factor $q$ [$\gamma$] 
which is irrelevant for the application, exactly the Lorentz transformation of 
the year 1904'' \cite{31}.

Are Voigt transformations and Lorentz transformations indeed equivalent? The 
answer depends on the reading of Voigt transformations: modern reader can 
see quite different contexts for them compared to Voigt's contemporaries. 
At first sight, the scale factor ``indicates only a uniform magnification of 
the scales of space and time, or what is the same thing, a change of units.
It does not introduce any essential modification'' \cite{34}. The same 
argument is also given in \cite{35}.

Indeed, let us assume that to define a time unit the observer in the moving
frame uses a standard atomic clock of the {\ae}ther frame. As this clock moves
with respect to the observer, his inferred time unit will be $\gamma$ times
larger than if he would use, as tacitly assumed in usual Lorentz 
transformations, an atomic clock at rest in his own frame as the standard 
clock. Modern definition of the length unit relates it through the light 
velocity $c$ to the time unit. Therefore, the length unit will be also 
$\gamma$ times larger, and to express the Lorentz transformations in these new
units of the moving frame we should divide primed coordinates by $\gamma$.
As a result, we get precisely the Voigt transformations (\ref{eq17}).

Therefore, the relativistic reading of Voigt transformations is certainly 
possible and the resulting theory will be completely equivalent (although
perhaps less convenient) to special relativity. Note that the introduction of 
the preferred {\ae}ther frame in this case is purely formal: we can choose any 
inertial reference frame instead of the {\ae}ther frame and use its 
standard clocks to define time units in other inertial reference frames.

However, surprisingly enough, at will a non-relativistic Galilean 
reading of Voigt transformations (\ref{eq17}) is also possible. Let us 
consider a hypothetical Newtonian world with the truly {\ae}ther frame $S$ 
which is the only inertial frame in which light propagates isotropically with 
velocity $c$. Transition to an other inertial frame $S^\prime$ is given
by Galilean transformations
\begin{equation}
\begin{array}{l}
\xi=x-V\, t, \\
\eta=y, \\
\zeta=z, \\
\tau=t.
\end{array}
\label{eq19}
\end{equation} 
This form of Galilean transformations assumes Newtonian absolute time. 
However, if the velocity $V$ of $S^\prime$ with respect to the {\ae}ther frame 
$S$ is smaller than light velocity $c$, it is possible to perform 
Poincar\'{e}-Einstein synchronization of clocks in $S^\prime$ and as a result
we get another parametrization of the Galilean space-time in $S^\prime$.
Let us find this  parametrization \cite{36}.

Suppose a light signal is sent from the origin $O^\prime$ of the inertial
reference frame $S^\prime$ at absolute time $\tau_1$, it arrives to some 
point $P^\prime(\xi,\eta,\zeta)$ at absolute time $\tau_2$, is instantaneously 
reflected by a mirror at $P^\prime$ and returns to $O^\prime$  at absolute 
time $\tau_3$. According to Poincar\'{e}-Einstein synchronization convention,
to synchronize the clock at $P^\prime$ with the clock at $O^\prime$,
one should set its reading to $t^\prime_2=(\tau_1+\tau_3)/2$ at an instant of
signal reflection. Taking
\begin{equation}
\tau_3=\tau,\;\;\; \tau_1=\tau-\frac{r^\prime}{c_1^\prime}-
\frac{r^\prime}{c_2^\prime},
\label{eq20}
\end{equation} 
where $r^\prime=\sqrt{\xi^2+\eta^2+\zeta^2}$, and $c_1^\prime,c_2^\prime$
are light velocities in the frame $S^\prime$ when it moves from $O^\prime$
to $P^\prime$ and back respectively, we get
\begin{equation}
t^\prime_2=\tau-\frac{1}{2}\left (\frac{r^\prime}{c_1^\prime}+
\frac{r^\prime}{c_2^\prime}\right ).
\label{eq21}
\end{equation}
Therefore, at the time $\tau_3=\tau$ at $O^\prime$, the Poincar\'{e}-Einstein 
synchronized clock at $P^\prime$ will show
\begin{equation}
t^\prime=t^\prime_2+\frac{r^\prime}{c_2^\prime}=\tau-\frac{1}{2}\left (
\frac{r^\prime}{c_1^\prime}-\frac{r^\prime}{c_2^\prime}\right )=
\tau-\frac{r^\prime}{2}\,\frac{c_2^\prime-c_1^\prime}{c_1^\prime c_2^\prime}.
\label{eq22}
\end{equation}
On the other hand, according to the Galilean velocity addition formula,
\begin{eqnarray} &&
c^2=(\vec{c}_1^\prime+\vec{V})^2=c_1^{\prime\,2}+2c_1^\prime\,V
\cos{\theta^\prime}+V^2,\nonumber \\ &&
c^2=(\vec{c}_2^\prime+\vec{V})^2=c_2^{\prime\,2}-2c_2^\prime\,V
\cos{\theta^\prime}+V^2,
\label{eq23}
\end{eqnarray}
where $\theta^\prime$ is the angle between the radius-vector $\vec{r}^\prime$
and the $x^\prime$-axis. The positive solutions of (\ref{eq23}) are
\begin{eqnarray} &&
c_1^\prime=-V\cos{\theta^\prime}+\sqrt{V^2\cos^2{\theta^\prime}+c^2-V^2},
\nonumber \\ &&
c_2^\prime=V\cos{\theta^\prime}+\sqrt{V^2\cos^2{\theta^\prime}+c^2-V^2}.
\label{eq24}
\end{eqnarray}
Substituting these velocities into (\ref{eq22}), we get (instead of 
$(\xi,\eta,\zeta)$, we have returned to the more traditional 
$(x^\prime,y^\prime, z^\prime)$ notation)
\begin{equation}
t^\prime=\tau-\frac{r^\prime\, V\,\cos{\theta^\prime}}{c^2-V^2}=
\tau-\gamma^2\,\frac{V}{c^2}\,x^\prime.
\label{eq25}
\end{equation}
Therefore, if the Poincar\'{e}-Einstein synchronization convention is adopted,
instead of Galilean transformations we will get Zahar transformations  
\cite{36}:
\begin{eqnarray} &&
x^\prime=x-V\,t, \nonumber \\ &&
y^\prime=y,\nonumber \\ &&
z^\prime=z,\nonumber \\ &&
t^\prime=t-\gamma^2\,\frac{V}{c^2}\,x^\prime=\gamma^2\left (t-\frac{V}{c^2}x
\right).
\label{eq26}
\end{eqnarray}
Let us note that, contrary to a prevailing belief, the non-relativistic limit
of Lorentz transformations, when $\beta=V/c\ll 1$, is not the Galilean 
transformations (\ref{eq19}) but the Zahar transformations (\ref{eq26})
in which $\beta^2$ terms are neglected \cite{37A,37,38}, so that 
$\gamma^2\approx 1$. The Galilean limit additionally requires
\begin{equation}
\beta\ll\frac{ct}{x},\;\;\;\frac{x}{ct}\sim\beta,
\label{eq27}
\end{equation}
which is not necessarily true if the spatial separation $x$ between two events
is comparable or larger than temporal separation $ct$. It seems that this 
simple fact is not widely known as witnessed by its rediscovery in 
\cite{39,40}. The reason for this mismatch of non relativistic limits is the 
use of different synchronization conventions in Lorentz transformations and
Galilean transformations: for sufficiently distant events in Minkowski world
one cannot ignore not absolute nature of distant simultaneity and only when
Poincar\'{e}-Einstein synchronization convention is adopted in Galilean world
too the resulting Zahar transformations share the common non relativistic limit
with the Lorentz transformations \cite{39}.

Now let us assume that inhabitants of the frame $S^\prime$ decided to use
an atomic clock of the frame $S$ as the time standard (and let us assume that
clocks when in motion behave as prescribed by Newtonian physics of Galilean
space-time). Then (\ref{eq26}) shows that the time unit so defined in 
$S^\prime$ will be $\gamma^2$ times larger than the time unit inferred from
the stationary atomic clock in $S^\prime$, as assumed in (\ref{eq26}).
Length units, as usually, are determined from the time unit and light 
velocity. However we should be careful here because the light velocity is not 
isotropic in $S^\prime$ in Galilean world even under the 
Poincar\'{e}-Einstein synchronization convention, if the ``natural'' units,
as in (\ref{eq26}), are used. The velocity addition rule that follows from 
(\ref{eq26}) has the form
\begin{equation}
V_x^\prime=\frac{V_x-V}{\gamma^2\left(1-\frac{V}{c^2}V_x\right)},\;\;
V_y^\prime=\frac{V_y}{\gamma^2\left(1-\frac{V}{c^2}V_x\right)},\;\;
V_z^\prime=\frac{V_z}{\gamma^2\left(1-\frac{V}{c^2}V_x\right)}.
\label{eq28}
\end{equation}
Suppose light signal moves in $S^\prime$ along the $x$-axis with velocity 
$c_\parallel=V_x^\prime$. Then (\ref{eq28}) shows that, in the frame $S$,
$V_y=V_z=0$. Hence $V_x=c$ and
\begin{equation}
c_\parallel=\frac{c-V}{\gamma^2\left(1-\frac{V}{c}\right)}=\frac{c}{\gamma^2}.
\label{eq29}
\end{equation}
If the light impulse in $S^\prime$ moves along the $y$-axis with velocity
$c_\perp=V_y^\prime$, then (\ref{eq28}) shows that $V_x=V,\,V_z=0$, and hence
$V_y=\sqrt{c^2-V^2}$. In this case (\ref{eq28}) will give
\begin{equation}
c_\perp=\frac{\sqrt{c^2-V^2}}{\gamma^2\left(1-\frac{V^2}{c^2}\right)}=
\frac{c}{\gamma}.
\label{eq30}
\end{equation}
Therefore, when the time unit is increased $\gamma^2$ times, the length unit
in the $x$-direction (parallel to the $S^\prime$ velocity) will not change 
at all and in the transverse $y$- and $z$-directions will be increased 
$\gamma$ times. To write Zahar transformations (\ref{eq26}) in these new 
units, we must therefore divide $t^\prime$ by $\gamma^2$, $y^\prime$ and 
$z^\prime$ by $\gamma$, and leave $x^\prime$ unchanged. As a result we again 
obtain Voigt transformations (\ref{eq17}). 

As we see, Voigt transformations by themselves are empirically empty
without detailing how coordinates involved in these transformations are 
related to the behavior of real clocks and rulers. Voigt transformations
require an additional input about the nature of real clocks not to be just 
a mathematical exercise but an instrument to answer questions about the physics
of the real world . Let us demonstrate this by considering a natural question 
whether the Poincar\'{e}-Einstein clock synchronization is equivalent 
to another synchronization method by slow clock transport.

Let us consider an inertial reference frame $S^\prime$ moving with respect to 
{\ae}ther frame $S$ (which is singled out by Voigt transformations) with 
velocity $V$. Clocks in $S^\prime$ are synchronized by Poincar\'{e}-Einstein 
procedure. Let one of the so synchronized clocks be transported slowly in 
$S^\prime$ from one point $A$ to an another point $B$ with the velocity 
$u^\prime\ll c$ along the $x$-axis (for simplicity). If this process takes 
time $\tau$ according to the clocks in $S^\prime$, then $\Delta \xi=u^\prime
\tau$ and the inverse Voigt transformations (\ref{eq18}) will give
\begin{equation}
\Delta x=\gamma_V^2(u^\prime+V)\tau,\;\;\;\Delta t=\gamma_V^2\left [1+
\frac{u^\prime V}{c^2}
\right ] \tau.
\label{eq31}
\end{equation}
The Voigt transformations (\ref{eq17}) imply just the same velocity addition 
rule as the Lorentz transformations. Therefore, with respect to $S$, the clock 
moving in $S^\prime$ has the velocity
\begin{equation}
u=\frac{u^\prime+V}{1+\frac{V}{c^2}u^\prime}\approx (u^\prime+V)\left (
1-\frac{V}{c^2}u^\prime\right )\approx V+\frac{u^\prime}{\gamma_V^2}.
\label{eq32}
\end{equation}
Hence, according to (\ref{eq17}) and (\ref{eq31}), while moving from $A$ to 
$B$, the clock will measure the proper time
\begin{equation}
\tau^\prime=\Delta t-\frac{u}{c^2}\Delta x\approx\gamma_V^2\tau\left [1-
\frac{V^2}{c^2}-\frac{u^\prime V}{c^2\gamma_V^2}\right ]=
\left (1-\frac{u^\prime V}{c^2}\right )\tau.
\label{eq33}
\end{equation}
However, this time interval is measured in time units of the reference frame
moving with velocity $u$, not $V$. Therefore, we need to rescale it in  
the $S^\prime$ time units in order to compare to the reading of the clock 
at rest in  $S^\prime$ at point $B$. This rescaling cannot be done on
the base of Voigt transformations alone and this is the point where an extra
empirical input about the behavior of real clocks is needed. In particular 
we must remember how time units for Voigt transformations were defined if
we stick the interpretation of these transformations given above with either
Lorentz or Zahar transformations as the primary source describing the 
behavior of real clocks. Having all this in mind, we write the rescaling as
follows
\begin{equation}
\tilde \tau=\tau^\prime\,\frac{\gamma_u^n}{\gamma_V^n},
\label{eq34}
\end{equation}
where $n=1$, if the real clocks behave as prescribed by special relativity,
and $n=2$, if the real clocks are Galilean. But from (\ref{eq32}) it is easy
to find that
\begin{equation}
\gamma_u\approx \gamma_V\left( 1+\frac{u^\prime V}{c^2}\right ).
\label{eq35}
\end{equation}
Therefore, from (\ref{eq33}) and (\ref{eq35}), we get
\begin{equation}
\tilde \tau= \tau\left (1-\frac{u^\prime V}{c^2}\right )
\left (1+n\frac{u^\prime V}{c^2}\right )\approx \tau\left [1+(n-1)
\frac{u^\prime V}{c^2}\right ].
\label{eq36}
\end{equation}
As we see, $\tilde \tau=\tau$ if and only if $n=1$. That is only in the 
Minkowski world the Poincar\'{e}-Einstein synchronization and 
synchronization by slow clock transport agree with each other. 
In Galilean world these two synchronization methods will give different 
results. 

Robertson \cite{41} and later Mansouri and Sexl \cite{42} developed a general
framework for test theories of special relativity which can be considered
as a generalization of Voigt, Lorentz and Zahar transformations. These
generalized transformations can be written in the form \cite{43}
\begin{eqnarray} &&
x^\prime=\lambda\,\frac{c_\parallel}{c_\perp}\,\gamma(x-V\,t), \nonumber \\ &&
y^\prime=\lambda y,\nonumber \\ &&
z^\prime=\lambda z,\nonumber \\ &&
t^\prime=\lambda\,\frac{c}{c_\perp}\,\gamma\left (t-\frac{V}{c^2}x
\right),
\label{eq37}
\end{eqnarray}
where $c_\parallel$ and $c_\perp$ are two-way  light speeds in the moving 
frame $S^\prime$ in the parallel and perpendicular directions to the
the velocity $\vec{V}$ of $S^\prime$ with respect to the preferred  
({\ae}ther) frame $S$, in which light propagates isotropically with the speed 
$c$. The parameters $c_\parallel/c$, $c_\perp/c$ and the conformal factor 
$\lambda$ are, in general, functions of $\gamma$. For example, if 
$\lambda=1,\,c_\parallel=c_\perp=c$, we get from (\ref{eq37}) Lorentz 
transformations; if $\lambda=1,\,c_\parallel=c/\gamma^2,\,c_\perp=c/\gamma$, 
we get Zahar transformations, and, finally, if $\lambda=\gamma^{-1},\,
c_\parallel=c_\perp=c$, we get Voigt transformations. It 
can be shown using (\ref{eq37}) that ``clock synchronization by clock 
transport and by the Einstein procedure agree if and only if the time 
dilatation factor is given exactly by the special relativistic value'' 
\cite{42}, that is if $\lambda\,c/c_\perp=1$. 

We hope that our above discussions of the Voigt transformations convinced you 
that a modern reader can find a lot of relativistic content in these 
transformations. However, all of them are later time readings and we suspect
that none were actually possible at Voigt's time. It would be not true to
say that Voigt was at the origin of relativistic revolution. Alas, he missed
this opportunity. 
 
\section{Another derivation of Voigt transformations}
One of the reasons why Voigt's 1887 paper remained unnoticed was its clumsy 
and unwieldy derivation of Voigt's transformations. In this section we will try
to give more modern and succinct derivation which will reveal basic physical
inputs and assumptions behind it. Inspiration for this derivation and its 
subsequent generalization comes mostly from one postulate derivation of 
Lorentz transformations originated in 1910 by von Ignatowsky \cite{44,44AA} 
\footnote{An English translation of \cite{44} is available at 
\newline {\scriptsize https://en.wikisource.org/wiki/Translation:Some\_General\_Remarks\_on\_the\_Relativity\_Principle}} and further 
developed by Frank and Rothe \cite{44A}. More precisely, our approach rests 
heavily on the ideas presented in \cite{45}. Eventually we generalize the 
approach of \cite{45} by relaxing  the requirement of spatial isotropy.

Under the assumption that measuring rods do not change their length when 
gently set into a state of uniform motion, the Galilean transformations 
(\ref{eq19}) are simply a statement that the length of a finite interval is
an additive quantity: the length of a union of two intervals is equal to the 
sum of their lengths. However, it seems not too fantastic to imagine that Voigt
in 1887 could question whether the measuring rods really do not change their 
lengths when moving through {\ae}ther. At least in 1888 Oliver Heaviside 
showed that the electric field of a charge in motion relative to the {\ae}ther 
is no longer spherically symmetric and becomes distorted in the the 
longitudinal direction. Influenced by the Heaviside's result, FitzGerald and 
later Lorentz came out with a hypothesis that Heaviside distortion might be 
applied to the molecular forces too and thus rigid bodies when moving through 
the {\ae}ther should experience deformations. In particular, in 1889 
FitzGerald published in then little known American journal {\it Science} 
a letter in which he wrote:

``I have read with much interest Messrs. Michelson and Morley's wonderfully 
delicate experiment attempting to decide the important question as to how far 
the {\ae}ther is carried along by the earth. Their result seems opposed to 
other experiments showing that the {\ae}ther in the air can be carried along 
only to an inappreciable extent. I would suggest that almost the only 
hypothesis that can reconcile this opposition is that the length of material 
bodies changes, according as they are moving through the {\ae}ther or across 
it, by an amount depending on the square of the ratio of their velocities to 
that of light. We know that electric forces are affected by the motion of the 
electrified bodies relative to the {\ae}ther, and it seems a not improbable 
supposition that the molecular forces are affected by the motion, and that the 
size of a body alters consequently'' \cite{46}. 

Interestingly, this FitzGerald's publication sank into oblivion and was 
virtually unknown until 1967 when it was unearthed by Brush \cite{46}. 
Paradoxically, FitzGerald himself was not aware whether his letter to 
{\it Science} was published or not: ``FitzGerald did publish the contraction 
hypothesis before Lorentz, but no one involved in later research on the subject
knew about this publication, including FitzGerald himself'' \cite{46}. He did 
promote his deformation idea mostly in lectures and was relieved to know that
in 1892 Lorentz came out with essentially the same idea. In 1894 FitzGerald
wrote to Lorentz: ``I am particularly delighted to hear that you agree with me,
for I have been rather laughed at for my view over here'' \cite{46}.

If we accept the possibility that measuring rods can change their length when
in motion, instead of $x=x^\prime+Vt$ we should write $x=k_1(V)x^\prime+Vt$,
where $k_1(V)$ accounts for a possible change of the measuring rod's length
in the moving frame $S^\prime$. But from the point of view of an observer in
$S^\prime$, the reference frame $S$ moves with velocity $-V$ (although there
are some subtleties in this reciprocity principle \cite{47,48}, we assume that
in 1887 Voigt probably would not question it). Therefore we can analogously
write $x^\prime=k_2(-V)x-Vt^\prime$. If we assume the validity of the 
relativity principle, then $k_1(V)=k_2(V)$, and if we assume the spatial 
isotropy, then $k_1(-V)=k_1(V)$. However, for Voigt, a true believer in 
{\ae}ther, relativity principle was not evident, so we do not 
assume its validity for a while, neither the spatial isotropy. Hence, we have
\begin{equation}
x^\prime=\frac{1}{k_1(V)}\left (x-Vt\right ),\;\;\;
x=\frac{1}{k_2(-V)}\left (x^\prime+Vt^\prime\right ).
\label{eq38}
\end{equation}
From these equations we can express $t^\prime$ in terms of $x$ and $t$,
or $t$ in terms of $x^\prime$ and $t^\prime$:
\begin{equation}
t^\prime=\frac{1}{k_1(V)}\left [t-\frac{1-K(V)}{V}\,x\right],\;\;\;
t=\frac{1}{k_2(-V)}\left [t^\prime+\frac{1-K(V)}{V}\,x^\prime\right],
\label{eq39}
\end{equation}
where $K(V)=k_1(V)k_2(-V)$. Therefore, we get the transformation
\begin{eqnarray} &&
x^\prime=\frac{1}{k_1(V)}\,(x-V\,t), \nonumber \\ &&
y^\prime=y,\nonumber \\ &&
z^\prime=z,\nonumber \\ &&
t^\prime=\frac{1}{k_1(V)}\,\left [t-\frac{1-K(V)}{V}\,x\right ],
\label{eq40}
\end{eqnarray}
and its inverse
\begin{eqnarray} &&
x=\frac{1}{k_2(-V)}\,(x^\prime+V\,t^\prime), \nonumber \\ &&
y=y^\prime,\nonumber \\ &&
z=z^\prime,\nonumber \\ &&
t=\frac{1}{k_2(-V)}\,\left [t^\prime+\frac{1-K(V)}{V}\,x^\prime\right ].
\label{eq41}
\end{eqnarray}
We have assumed that the transverse coordinates do not change. This is not
the most general possibility but probably the one which Voigt's contemporaries
would infer from  Heaviside's result on the moving Coulomb field distortions.

Voigt was interested in wave equation (\ref{eq1}) which is obviously invariant
under the scale transformations. Considering all coordinates that differ only
by a scale change as equivalent, we can choose new coordinates
\begin{equation}
X^\prime=k_1(V)x^\prime, \;\;Y^\prime=k_1(V)y^\prime,\;\;
Z^\prime=k_1(V)z^\prime, \;\;T^\prime=k_1(V)t^\prime,
\label{eq42}
\end{equation}
to regain the Galilean relation $x=X^\prime+Vt$. In new coordinates 
(\ref{eq40}) takes the form
\begin{eqnarray} &&
X^\prime=x-V\,t, \nonumber \\ &&
Y^\prime=k_1(V)\,y,\nonumber \\ &&
Z^\prime=k_1(V)\,z,\nonumber \\ &&
T^\prime=t-\frac{1-K(V)}{V}\,x,
\label{eq43}
\end{eqnarray}
while its inverse (\ref{eq41}) transforms into
\begin{eqnarray} &&
x=\frac{1}{K(V)}\left[X^\prime+V\,T^\prime\right],, \nonumber \\ &&
y=k^{-1}_1(V)\,Y^\prime,\nonumber \\ &&
z=k^{-1}_1(V)\,Z^\prime,\nonumber \\ &&
t=\frac{1}{K(V)}\left[T^\prime+\frac{1-K(V)}{V}\,X^\prime \right ].
\label{eq44}
\end{eqnarray}
The velocity addition rule following from (\ref{eq44}) has the form
\begin{eqnarray} &&
v_x=\frac{v^\prime_x+V}{1+\frac{1-K(V)}{V}\,v^\prime_x},\nonumber \\ &&
v_y=\frac{K(V)\,v^\prime_y}{k_1(V)\left[1+\frac{1-K(V)}{V}\,v^\prime_x\right]},
\;\;\;
v_z=\frac{K(V)\,v^\prime_z}{k_1(V)\left[1+\frac{1-K(V)}{V}\,v^\prime_x\right]}.
\label{eq45}
\end{eqnarray}
If we now require the universality of the light velocity as it follows from 
the assumed invariance of the wave equation, and assume $v^\prime_x=c$,
$v^\prime_y=v^\prime_z=0$, then we get from (\ref{eq45}) $v_y=v_z=0$,
\begin{equation}
c=v_x=\sqrt{c^2-v_y^2+v_z^2}=\frac{c+V}{1+\frac{1-K(V)}{V}\,c},
\label{eq46}
\end{equation}
and therefore,
\begin{equation}
K(V)=1-\frac{V^2}{c^2}\equiv \frac{1}{\gamma^2}.
\label{eq47}
\end{equation}
On the other hand, if we take  $v^\prime_y=c$, $v^\prime_x=v^\prime_z=0$, from
(\ref{eq45}) we obtain
\begin{equation}
v_x=V,\;\;\;v_y=\frac{K(V)\,c}{k_1(V)}=\frac{c}{\gamma^2\,k_1(V)},\;\;
v_z=0,
\label{eq48}
\end{equation}
which together with the condition $v_x^2+v_y^2+v_z^2=c^2$ then determines 
$k_1(V)$:
\begin{equation}
k_1(V)=\frac{1}{\gamma}.
\label{eq49}
\end{equation}
Substituting (\ref{eq47}) and (\ref{eq49}) into (\ref{eq43}) we get indeed
the Voigt transformations (\ref{eq17}).

Now we will show that a little generalization of this derivation provides an 
interesting perspective on the relativity principle. Let us return to the 
original coordinates before rescaling (\ref{eq42}), but this time we will 
not assume that the transverse coordinates remain unchanged. So we modify 
(\ref{eq40}) and (\ref{eq41}) into
\begin{eqnarray} &&
x^\prime=\frac{1}{k_1(V)}\,(x-V\,t), \nonumber \\ &&
y^\prime=\lambda(V)y,\nonumber \\ &&
z^\prime=\lambda(V)z,\nonumber \\ &&
t^\prime=\frac{1}{k_1(V)}\,\left [t-\frac{1-K(V)}{V}\,x\right ]
\label{eq50}
\end{eqnarray}
and 
\begin{eqnarray} &&
x=\frac{1}{k_2(-V)}\,(x^\prime+V\,t^\prime), \nonumber \\ &&
y=\lambda^{-1}(V)y^\prime,\nonumber \\ &&
z=\lambda^{-1}(V)z^\prime,\nonumber \\ &&
t=\frac{1}{k_2(-V)}\,\left [t^\prime+\frac{1-K(V)}{V}\,x^\prime\right ]
\label{eq51}
\end{eqnarray}
respectively. From (\ref{eq51}) we get the following velocity addition rule:
\begin{eqnarray} &&
v_x=\frac{v^\prime_x+V}{1+\frac{1-K(V)}{V}\,v^\prime_x}=F(v^\prime_x,V),
\nonumber \\ &&
v_y=\frac{k_2(-V)\,v^\prime_y}{\lambda(V)\left[1+\frac{1-K(V)}{V}\,v^\prime_x
\right]},\;\;\;
v_z=\frac{k_2(-V)\,v^\prime_z}{\lambda(V)\left[1+\frac{1-K(V)}{V}\,v^\prime_x
\right]}.
\label{eq52}
\end{eqnarray}
The reasoning analogous to what followed after (\ref{eq45}) makes it clear that 
the universality of the light velocity will then demand the validity of 
(\ref{eq47}) and
\begin{equation}
\frac{k_2(-V)}{\lambda(V)}=\frac{1}{\gamma}.
\label{eq53}
\end{equation}
From (\ref{eq47}) and (\ref{eq53}) we can get
\begin{equation}
\lambda(V)=\frac{k_2(-V)}{\sqrt{K(V)}}=\sqrt{\frac{k_2(-V)}{k_1(V)}},
\label{eq54}
\end{equation}
and
\begin{equation}
\frac{1}{k_1(V)}=\sqrt{\frac{k_2(-V)}{k_1(V)}}\,\frac{1}{\sqrt{k_1(V)k_2(-V)}}
=\frac{\lambda(V)}{\sqrt{K(V)}}=\lambda(V)\gamma.
\label{eq55}
\end{equation}
Therefore, (\ref{eq50}) takes the form
\begin{eqnarray} &&
x^\prime=\lambda(V)\gamma\left(x-V\,t\right ), \nonumber \\ &&
y^\prime=\lambda(V)y,\nonumber \\ &&
z^\prime=\lambda(V)z,\nonumber \\ &&
t^\prime=\lambda(V)\gamma\,\left (t-\frac{V}{c^2}\,x\right ).
\label{eq56}
\end{eqnarray}
Both Einstein \cite{1} and Poincar\'{e} \cite{15,16} got these 
$\lambda$-Lorentz transformations and then both argued that $\lambda(V)=1$.
The arguments of Einstein, in section 3 of his paper, were more physical as 
he correctly identified the physical meaning of $\lambda(V)$: it corresponds 
to a contraction of the
length of a moving rod which is perpendicular to the direction of its motion.
''He makes then the clever statement that, because the rod is perpendicular to 
the direction of motion, this possible contraction can well depend on the 
relative velocity, but not on its sign!'' \cite{17}. Therefore, $\lambda(-V)=
\lambda(V)$. On the other hand, if we adopt the relativity principle so that
the {\ae}ther frame loses its privileged role, then transformations 
(\ref{eq56}) should form a group which implies
\begin{equation}
\lambda(V_1\oplus V_2)=\lambda(V_1)\lambda(V_2),
\label{eq57}
\end{equation}
where $V_1\oplus V_2$ denotes the relativistic sum of velocities. In 
particular, if we take $V_1=-V_2=V$, then we get $\lambda(V)\lambda(-V)=
\lambda(0)=1$ which in combination with $\lambda(-V)=\lambda(V)$ and 
positivity of $\lambda(V)$ implies $\lambda(V)=1$.

Poincar\'{e}'s reasoning,  in section 4 of his paper, was more formal 
and mathematical. He makes a thorough 
analysis of the Lorentz's group including not only the boosts, but also the
spatial rotations and gets $\lambda(V)=1$ essentially from the following group 
theoretical argument (for more complete analysis of Poincar\'{e}'s derivation,
see \cite{49A,49B,49C}). Let us denote the $\lambda$-Lorentz transformations
(\ref{eq56}) by $\hat L$ and let $\hat R=\hat R^{-1}$ be a rotation by $\pi$
around the $y$-axis which changes signs of the $x$ and $z$ coordinates and 
leaves $y$ and $t$ unchanged. Then by the group property $\hat R\hat L\hat R$
must belong to the same Lorentz group. But, as it can be easily checked,
\begin{eqnarray} &&
\hat R\hat L\hat R(x)=\lambda(V)\gamma\left(x+V\,t\right ), \nonumber \\ &&
\hat R\hat L\hat R(y)=\lambda(V)y,\nonumber \\ &&
\hat R\hat L\hat R(z)=\lambda(V)z,\nonumber \\ &&
\hat R\hat L\hat R(t)=\lambda(V)\gamma\,\left (t+\frac{V}{c^2}\,x\right ).
\label{eq56A}
\end{eqnarray}
Comparing with (\ref{eq56}), we conclude that this is a is pure boost with 
velocity $-V$ instead of $V$ and, as the boost is uniquely determined by the
velocity, we must have $\lambda(-V)=\lambda(V)$, which in combination with
$\lambda(V)\lambda(-V)=1$ and positivity of $\lambda(V)$ implies 
$\lambda(V)=1$. 

Note that the multiplication law (\ref{eq57}) first appeared in Poincar\'{e}'s 
May 1905 letter to Lorentz \cite{49}. Einstein in \cite{1} doesn't provide 
this multiplication law in its full generality --- he considers only mutually 
inverse transformations with $V_1=-V_2$.

In fact, both derivations of $\lambda(V)=1$ implicitly assume spatial 
isotropy. Although natural, this is not the most general possibility. So let
us go ahead without this assumption and see what gems were thrown overboard
by Einstein and Poincar\'{e} when insisting on it.

First of all, Poincar\'{e} missed an opportunity to discover that Maxwell 
equations and wave equation are invariant not only with respect to the Lorentz
group but also with respect to much wider class of conformal transformations
(we think, for Einstein this feat was unrealistic in 1905). This fact was
established several years later, namely in 1910, by Cunningham and Bateman 
\cite{50} (their road to this discovery is described in \cite{50A}). Conformal 
group is a fifteen parameter Lie group. It contains as a subgroup the ten 
parameter Poincar\'{e} group (inhomogeneous Lorentz transformations). A general 
Poincar\'{e} transformation is a combination of a rotation (three parameters)
with a boost (three parameters), followed by a translation in space-time
(four  parameters). Additional five parameters of the conformal group are:
$\lambda$ of the scale transformation, promoted to an independent parameter
representing dilatation, and four parameters of the so called special 
conformal transformation ($x_0=x^0=ct,x_1=-x^1=-x,x_2=-x^2=-y,x_3=-x^3=-z$)
\begin{equation}
x^{\prime\,\mu}=\frac{x^\mu-\alpha^\mu \,x_\nu x^\nu}{1-2\alpha_\mu x^\mu+
\alpha_\mu \alpha^\mu\,x_\nu x^\nu}.
\label{eq58}
\end{equation}
Taking $\alpha^\mu=(0,\vec{a})$, in the non-relativistic limit $c\to\infty,\,
|\vec{a}|\to 0,\,c^2\vec{a}\to\vec{g}/2$, where $\vec{g}$ is some constant 
3-vector, we get
\begin{equation}
t^\prime=t,\;\;\;\vec{r}{\,^\prime}=\vec{r}-\frac{\vec{g}\,t^2}{2}.
\label{eq59}
\end{equation}
Therefore, in the non-relativistic limit the transformation (\ref{eq58}) 
describes the transition to a non-inertial frame moving with a constant 
acceleration $\vec{g}$.

Special conformal transformation is tightly related to uniformly accelerated 
motion in relativistic case too (see the appendix \ref{AppA}). 
It seems therefore tempting to interpret the special conformal transformation
(\ref{eq58}) as a transition to a uniformly accelerated reference frame.
There were long-lasting efforts \cite{51,52,53,54} to make sense of this
interpretation however they were not very successful, and the modern 
consensus is that such interpretation is untenable \cite{55,56}. Passive
interpretation (as a change of the reference frame) is not always  
advantageous whereas an active interpretation of transformations and the 
corresponding notion of symmetries is often more fruitful.

The structure of the conformal group (more precisely, the commutation 
relations of its Lie algebra) indicates that ``a Poincar\'{e} invariant 
theory that is also conformally invariant necessarily enjoys scale 
invariance'', but the contrary is not true: ``group theory does not establish 
conformal invariance from scale and Poincar\'{e} invariance'' \cite{57}.
This fact probably explains why Poincar\'{e} missed an opportunity to 
discover conformal invariance of the Maxwell equations (for a pedagogical
exposition of the role of the conformal group in electrodynamics see, for 
example, \cite{57A}). Interestingly, if the dimensionality of space-time is 
greater than four, the Maxwell equations are no longer conformal invariant 
even though they remain scale invariant \cite{57}.   

The second postulate about the universality of light speed in inertial 
reference frames plays the role of founding principle in Einstein's version
of special relativity. One may wonder what is so special  about light that 
makes its speed a fundamental constant. In the Standard Model there is no
compelling theoretical reason to assume that the photon mass is strictly zero
\cite{58}. A tiny photon mass can be generated by means of the St\"{u}ckelberg 
mechanism that retains gauge invariance, unitarity and renormalizability
\cite{59}. Of course, if the Standard Model is a low energy remnant of a grand 
unification model based on a simple nonabelian group like $SU(5)$, $SO(10)$, 
or $E_6$, then the photon will be necessarily truly massless because one 
cannot generalize St\"{u}ckelberg mechanism to a non-Abelian gauge symmetry.
However, even in this case the masslessness of the photon originates from the
particular pattern of the electroweak symmetry breaking. One may suspect
therefore that the fundamental constant $c$ appearing in the Lorentz 
transformations is not necessarily light speed in its essence. This is indeed
the case as the following simple argument shows \cite{45}.
 
Let us return to the velocity addition rule (\ref{eq52}). If we assume the 
validity of the relativity principle, then $k_1(V)=k_2(V)$ and hence, $K(-V)=
K(V)$ so that the function $F(V_1,V_2)$ in (\ref{eq52}) is an odd function:
\begin{equation}
F(-V_1,-V_2)=-F(V_1,V_2).
\label{eq79}
\end{equation}
Let us combine this oddness of $F(V_1,V_2)$ with the reciprocity principle
$V_{AB}=-V_{BA}$ for the velocity $V_{AB}$ of an object $A$ relative to
an object $B$. We will have
\begin{eqnarray} &&
V_{AB}=F(V_{AC},V_{CB})=-V_{BA}=-F(V_{BC},V_{CA})=
\nonumber \\ &&
-F(-V_{CB},-V_{AC})=F(V_{CB},V_{AC}).
\label{eq80}
\end{eqnarray}
Therefore, $F(V_1,V_2)$ is a symmetric function of its arguments what in light 
of its definition in (\ref{eq52}) implies
\begin{equation}
\frac{1-K(V_2)}{V_2}\,V_1=\frac{1-K(V_1)}{V_1}\,V_2.
\label{eq81}
\end{equation}
This is only possible if
\begin{equation}
\frac{1-K(V)}{V^2}=k=\mathrm{const}.
\label{eq82}
\end{equation}
If $k<0$, then we will face a number of strange features like positive 
velocities that sum up in a negative velocity. Most importantly, in this
case it will be difficult if not impossible to define a causal structure
on the corresponding space-time \cite{45,47,60}.

If $k\ge 0$, we can take $k=1/c^2$, where $c$ is some constant with a 
dimensionality of velocity. Nothing however indicates in such a derivation 
that $c$ is the light velocity in vacuum. In this case $K(V)$ is given by
(\ref{eq47}) and the first equation in (\ref{eq52}) coincides to the 
relativistic addition law for parallel velocities. To promote $c$ to the
invariant velocity for perpendicular velocities too, we will need 
(\ref{eq53}). Then (\ref{eq52}) will completely coincide with the 
special-relativistic velocity-addition law. $k=0$ corresponds to the
$c\to \infty$ limit and gives Galilean velocity-addition law. Whether the 
light velocity in vacuum is the invariant velocity $c$ is the question about 
the properties of light and it has nothing to do with the logical structure of
special relativity. However, experimental limits on the nonzero photon mass
are very tough: the photon Compton wavelength is already known to be
at least comparable in dimension to the Earth-Sun distance \cite{61}. So
for all laboratory-scale purposes the photon mass safely may be taken as zero
and therefore $c$ may be considered as light velocity  for all practical 
purposes. 

\section{Lalan-Alway-Bogoslovsky transformations and anisotropic special 
relativity}
It was hinted in the previous section that $\lambda$-Lorentz 
transformations (\ref{eq56}) with $\lambda(V)$ satisfying the functional
equation (\ref{eq57}) are the most general transformations  that follow
from Einstein's two postulates under some natural assumptions (homogeneity 
of space and time, the reciprocity principle). To solve
the functional equation (\ref{eq57}) it is beneficial to realize, as it is 
well known,  that the natural parameter for special Lorentz transformations 
is not the velocity $V$ but the rapidity $\psi$ \cite{62} defined through
\begin{equation}
\tanh{\psi}=\beta=\frac{V}{c}.
\label{eq83}
\end{equation}
Rapidities are additive for the special Lorentz transformations, and the 
functional equation (\ref{eq57}) takes the form of the Cauchy exponential 
functional equation
\begin{equation}
\lambda(\psi_1+\psi_2)=\lambda(\psi_1)\,\lambda(\psi_2).
\label{eq84}
\end{equation}
Although there exist infinitely many wildly discontinuous solutions of
(\ref{eq84}), the continuous solutions, which are the only ones acceptable in
the context of $\lambda$-Lorentz transformations, all have the form \cite{63} 
\begin{equation}
\lambda(\psi)=e^{-b\psi}=\left(\frac{1-\beta}{1+\beta}\right )^{b/2}
\label{eq85}
\end{equation}
with some real constant $b$. Indeed,  we will have from (\ref{eq84}), 
$\lambda(\psi+d\psi)=\lambda(\psi)\lambda(d\psi)$, and because $\lambda(0)=1$,
$\lambda(\psi)+\lambda^\prime(\psi)d\psi\approx (1+\lambda^\prime(0)d\psi)
\lambda(\psi)$, which leads to the differential equation 
$\lambda^\prime(\psi)=\lambda^\prime(0)\lambda(\psi)$ with the solution 
(\ref{eq85}) with $\lambda^\prime(0)=-b$. 

Therefore, $\lambda$-Lorentz transformations (\ref{eq56}) take the form
\begin{eqnarray} &&
x^\prime=\left(\frac{1-\beta}{1+\beta}\right )^{b/2}
\gamma\left(x-V\,t\right ), \nonumber \\ &&
y^\prime=\left(\frac{1-\beta}{1+\beta}\right )^{b/2}y,\nonumber \\ &&
z^\prime=\left(\frac{1-\beta}{1+\beta}\right )^{b/2}z,\nonumber \\ &&
t^\prime=\left(\frac{1-\beta}{1+\beta}\right )^{b/2}
\gamma\,\left (t-\frac{V}{c^2}\,x\right ).
\label{eq86}
\end{eqnarray}
Let us introduce light-cone coordinates \cite{64}
\begin{equation}
u=ct+x,\;\;\;v=ct-x.
\label{eq87}
\end{equation}
Then we get from (\ref{eq86})
\begin{equation}
u^\prime=e^{-(1+b)\psi}\,u,\;\;\;v^\prime=e^{(1-b)\psi}\,v,\;\;\;
y^\prime=e^{-b\psi}\,y,\;\;\;z^\prime=e^{-b\psi}\,z.
\label{eq88}
\end{equation}
Using
\begin{equation}
\frac{u^\prime}{v^\prime}=e^{-2\psi}\,\frac{u}{v},\;\;\;
u^\prime\,v^\prime=e^{-2b\psi}u\,v,
\label{eq89}
\end{equation}
we can easily find invariant quantities
\begin{eqnarray} &&
\left(\frac{v^\prime}{u^\prime}\right )^b u^\prime v^\prime=
\left(\frac{v}{u}\right )^b u v,\;\;
\left(\frac{v^\prime}{u^\prime}\right )^b y^{\,\prime\,2}=
\left(\frac{v}{u}\right )^b y^2,\;\;
\left(\frac{v^\prime}{u^\prime}\right )^b z^{\,\prime\,2}=
\left(\frac{v}{u}\right )^b z^2,\nonumber \\ &&
\frac{u^\prime v^\prime}{y^{\,\prime\,2}}
=\frac{u v}{y^{2}},\;\;\frac{u^\prime v^\prime}{z^{\,\prime\,2}}
=\frac{u v}{z^{2}}.
\label{eq90}
\end{eqnarray}
Only first three of them are independent. Therefore, the following quantity, 
which was considered in the literature as a generalization of the relativistic 
interval, is invariant under $\lambda$-Lorentz transformations (\ref{eq56}):
\begin{eqnarray} &&
s^2=\left(\frac{v}{u}\right )^b\left (uv-y^2-z^2\right)\left(\frac{uv}
{uv-y^2-z^2}\right)^b=
\nonumber \\ &&
v^{2b}(uv-y^2-z^2)^{1-b}=\left(ct-x\right)^{2b}
\left(c^2t^2-x^2-y^2-z^2\right)^{1-b}.
\label{eq91}
\end{eqnarray}
Since the  $\lambda$-Lorentz transformations (\ref{eq56}) are linear, the 
differentials of coordinates are transformed as the coordinates themselves.
Consequently, the metric of the space-time, invariant with respect to 
the $\lambda$-Lorentz transformations (\ref{eq56}) has the form
\begin{equation}
ds^2=\left(c\,dt-dx\right)^{2b}
\left(c^2\,dt^2-dx^2-dy^2-dz^2\right)^{1-b}.
\label{eq92A}
\end{equation} 
Introducing  the fixed null-vector $n^\mu=(1,1,0,0)=(1,\vec{n}),\;
\vec{n}^{\,2}=1$, which defines a preferred null-direction in the space-time,
(\ref{eq92A}) can be rewritten as follows
\begin{equation}
ds^2=\left(n_\nu dx^\nu\right)^{2b}\left(dx_\mu dx^\mu\right)^{1-b},
\label{eq92}
\end{equation} 

As we see, in general, if the spatial isotropy is not assumed, Einstein's two 
postulates do not lead to Lorentz transformations and Minkowski metric.  
Instead they lead to more general $\lambda$-Lorentz transformations and 
Finslerian metric (\ref{eq92}). In the Finsler geometry the squared line 
element is homogeneous of second degree in the coordinate increments:
$ds^2=f(x_\mu,dx_\nu)$, $f(x_\mu,\alpha dx_\nu)=\alpha^2f(x_\mu,dx_\nu)$.
Although Finsler geometry is more general geometry than the Riemann geometry,
which is obtained than $f(x_\mu,dx_\nu)$ is a quadratic form in coordinate 
increments, almost all the results of Riemannian geometry can be developed in 
the Finsler case too \cite{65A}. For applications of Finsler geometry in 
physics see, for example, \cite{65,66}.

Note that the way how we obtained (\ref{eq92}) is highly non-unique. However
(\ref{eq92}) is the one of simplest Finsler metric which can be constructed
from the invariants of the $\lambda$-Lorentz transformations. That it may have 
a physical meaning is reinforced by the fact that (\ref{eq92}) emerges also 
in the process of deformation of the very special relativity symmetry group 
$ISIM(2)$ \cite{88}. 

$\lambda$-Lorentz transformations (\ref{eq56}) give the coordinate 
transformations between inertial frames in the case when one of them moves 
with respect to other with relative velocity $V$ along the $x$-axis 
(along the preferred direction $\vec{n}$). What about other inertial frames?

The corresponding generalized Lorentz transformations were obtained in 
\cite{70} by using  the Lobachevsky geometry of the 
special-relativistic velocity space. In the appendix \ref{AppB} we give 
a different derivation of these transformations. To summarize the results of 
this appendix, the Finsler metric (\ref{eq92}) is invariant under the 
generalized Lorentz transformations \cite{70}
\begin{equation}
x^{\prime\mu}=D(\lambda)R^\mu_{\;\nu}(\vec{m};\alpha)\,L^\nu_{\;\sigma}
(\vec{V})x^\sigma,
\label{eq114}
\end{equation}
where $D(\lambda)$, $R^\mu_{\,\nu}(\vec{m};\alpha)$ and $L^\nu_{\,\sigma}
(\vec{V})$ represent, respectively, the dilatation transformation, the spacial 
rotation and the Lorentz boost each defined in the appendix.

Using (\ref{eq103}) and (\ref{eq106}), we can get an explicit form of these
generalized Lorentz transformations:
\begin{eqnarray} &&
x_0^\prime=\left[\gamma(1-\vec{\beta}\cdot\vec{n})\right]^b\gamma\,(x_0-
\vec{\beta}\cdot\vec{r}), \nonumber \\ &&
\vec{r}^{\,\prime}=\left[\gamma(1-\vec{\beta}\cdot\vec{n})\right]^b\left\{
\vec{r}-\frac{\vec{\beta}\,(x_0-\vec{n}\cdot\vec{r})}{1-\vec{\beta}\cdot
\vec{n}}-
\right . \nonumber \\ && \left . 
\vec{n}\left[\gamma\,\vec{\beta}\cdot\vec{r}+\frac{\gamma-1}{\gamma}
\,\frac{\vec{n}\cdot\vec{r}}{1-\vec{\beta}\cdot\vec{n}}+\frac{(\gamma-1)
\vec{\beta}\cdot\vec{n}-\gamma\beta^2}{1-\vec{\beta}\cdot\vec{n}}\,x_0
\right]\right\}.
\label{eq115}
\end{eqnarray}
We have checked that these expressions are equivalent to  Eq.29 of \cite{70}.
If $\vec{\beta}$ and $\vec{n}$ are parallel and the $x$-axis is along 
$\vec{\beta}$, (\ref{eq115}) are reduced to the $\lambda$-Lorentz 
transformations (\ref{eq86}). Another interesting limiting case is when  
$\vec{\beta}$ and $\vec{n}$ are perpendicular to each other. Assuming that
the $x$ and $y$ axes are respectively along $\vec{\beta}$ and $\vec{n}$,
we get from (\ref{eq115})
\begin{eqnarray} &&
x^\prime=\gamma^b\left[x-Vt+\beta\,y\right],\nonumber \\ &&
y^\prime=\gamma^{1+b}\left[(1-\beta^2)\,y-\beta\,(x-Vt)\right ], 
\nonumber \\ &&
z^\prime=\gamma^b\,z,\nonumber \\ &&
t^\prime=\gamma^{1+b}\left[t-\frac{V}{c^2}\,x\right ].
\label{eq116}
\end{eqnarray}
It may appear that transformations (\ref{eq116}) violate the reciprocity 
principle because according to (\ref{eq116}) the origin $x=0,y=0,z=0$ of 
the frame $S$ moves in the frame $S^\prime$ with velocity $V^\prime_x=
-V/\gamma,\,V^\prime_y=\beta\,V,\,V^\prime_z=0$, not with $V^\prime_x=-V,\,
V^\prime_y=0,\,V^\prime_z=0$ as one might expect. However, remember that 
the generalized Lorentz transformations (\ref{eq114}) contain the rotation 
$R(\vec{m};\alpha)$. When passively interpreted, this rotation implies in 
the particular case of (\ref{eq116}) that the $x$ and $y$-axes of $S^\prime$ 
are rotated around the $z$-axis with an angle 
$\alpha$ such that $\sin{\alpha}=\beta,\,\cos{\alpha}=1/\gamma$ according to 
(\ref{eq98}) and (\ref{eq100}). No wonder that the vector $-\vec{V}$  has the 
coordinates $V^\prime_x=-V\cos{\alpha}=-V/\gamma$ and $V^\prime_y=
V\sin{\alpha}=\beta\,V$ in $S^\prime$.

To our best knowledge, it was Lalan \cite{71,72,73} who first derived 
$(1+1)$-dimensional $\lambda$-Lorentz transformations and  demonstrated that 
the metric invariant under such transformations was of pseudo-Finslerian type
rather than of the Minkowski pseudo-Riemannian. $(1+3)$-dimensional
generalized Lorentz transformations (\ref{eq115}) were first discovered by 
Alway \cite{74}. He gives no details on how the transformations were obtained 
but it seems that the spirit of the derivation was somewhat similar to the one 
presented above as he cites Pars \cite{75} and mentions that one of the Pars' 
assumptions, namely isotropic behavior of clocks, is not guaranteed in general.
Pars like von Ignatowsky \cite{44} provides the derivation of 
the Lorentz transformations without Einstein's second  postulate. He doesn't 
cite neither von Ignatowsky nor anybody else in his paper.  The fate of Pars's 
and von Ignatowsky's innovative derivations of Lorentz transformations were 
similar: ``like numerous others that followed, these have gone largely 
unnoticed'' \cite{76}.  Alway's generalized Lorentz transformations and 
the corresponding Finslerian space-time metric discovered by him also fell 
into oblivion immediately after it was recognized that experimental 
observations severely constrain spatial anisotropy essentially making the 
parameter $b$ zero \cite{77}. Fortunately, generalized Lorentz transformations 
(\ref{eq115}) were soon rediscovered by Bogoslovsky \cite{70,78} who then 
thoroughly investigated their physical consequences \cite{79,80,81}.
$(1+1)$-dimensional $\lambda$-Lorentz transformations for the case $b=1/2$
was independently rediscovered by Brown \cite{82} and then generalized to any
value of $b$ by Budden \cite{83}. Apparently, initially they were unaware of
Bogoslovsky's pioneering contributions. Later they cite Bogoslovsky in 
\cite{84,85} and \cite{86}, but neither  Lalan nor Alway are mentioned 
in their work. Slightly different perspective on Finslerian relativity is 
given by Asanov \cite{87}.

As was already mentioned, observations severely constrain the Finslerian 
parameter $b$. Some {\ae}ther-drift experiments imply $|b|<10^{-10}$,
while the Hug\-hes-Drever type limits on the anisotropy of inertia can 
potentially lower the limit on the Finslerian parameter $b$ up to
$|b|<10^{-26}$, though in a model-dependent way \cite{88}. It may appear
therefore that Strnad's conclusion \cite{77} that the special relativity is
valid for all practical purposes, and the Alway's extension of Lorentz 
transformations \cite{74} is superfluous, is well justified. However, 
notwithstanding our firm belief that special relativity is indeed valid for 
all practical purposes, its Finslerian extension described above, in our 
opinion, has a very significant conceptual value. Let us explain why. 

Besides the generalized pure Lorentz boosts (\ref{eq115}), the Finsler metric
(\ref{eq92}) is left invariant by rotations about preferred spatial direction 
$\vec{n}$ and by space-time translations $x^{\prime\mu}=x^\mu+a^\mu$. Together 
these transformations constitute an 8-parameter group of isometries of the 
Finsler space-time with metric (\ref{eq92}). This group is given a fancy name
DISIM$_b$(2) (deformed ISIM(2), $b$ being the deformation parameter) in 
\cite{88}. Both DISIM$_b$(2) and the 10-parameter Poincar\'{e} group are 
subgroups of the 11-parameter Weyl group (Poincar\'{e} transformations  
augmented with arbitrary dilations). It may appear therefore that DISIM$_b$(2) 
plays the role similar to the Poincar\'{e} group but it doesn't. The role
analogues to the Poincar\'{e} group is played by ISIM(2) which is the
$b\to 0$ limit (or better to say contraction in the sense of In\"{o}n\"{u} and 
Wigner \cite{89}) of DISIM$_b$(2). Although the Finsler metric (\ref{eq92})
is reduced to the Minkowski metric in the $b\to 0$ limit, the generalized 
Lorentz transformations (\ref{eq115}) are not reduced to the ordinary Lorentz 
transformations. Instead in this limit the transformations (\ref{eq115})
augmented with  rotations about the preferred direction $\vec{n}$ constitute
a 4-parameter subgroup of the Lorentz group called SIM(2) (similitude group of 
the plane consisting of dilations, translations and rotations of the plane).
If we further add space-time translations, we get a 8-parameter subgroup
of the Poincar\'{e} group ISIM(2) which is a semi-direct product of SIM(2)
with the group of space-time translations. Very special relativity developed 
by Cohen and Glashow \cite{90} assumes that the exact symmetry group of nature 
may be not the  Poincar\'{e} group but its subgroup ISIM(2).

It is known \cite{91,92} that the Lorentz group has no 5-parameter subgroup
and only one 4-parameter subgroup SIM(2) up to isomorphism. As a
result, very special relativity breaks Lorentz symmetry in a very mild and
minimal way. For amplitudes satisfying appropriate analyticity properties, 
$CPT$ discrete symmetry follows from SIM(2), but SIM(2) violates $P$ and
$T$ discrete symmetries \cite{90}. However, the incorporation of either $P$, 
$T$ or $CP$ discrete symmetries enlarges SIM(2) subgroup to the full Lorentz 
group and thus, for particle physics theories conserving any 
one of these three discrete symmetries Lorentz-violating effects in very 
special relativity are absent \cite{90}. The existence of the preferred 
light-like direction $n^\mu$ can be interpreted as the existence of 
light-like {\ae}ther. However, this peculiar form of {\ae}ther is very 
difficult to detect because it doesn't single out any preferred inertial 
reference frame and, since $CP$ violating effects are small, Lorentz-violating 
effects in very special relativity are expected to be also small.
 
Although the $b\to 0$ limits of (\ref{eq92}) and (\ref{eq115}) lead at the 
first sight naturally to the very special relativity, it is not so. When
$b=0$, or when the space is isotropic, we have no reason to introduce 
the preferred light-like direction $n^\mu$. Of course, we can do this 
artificially and consequently arrive to the generalized Lorentz 
transformations (\ref{eq115}) instead of usual  Lorentz transformations.
However, in this case $\vec{n}$ has no physical meaning and just serves to 
calibrate the orientations of space axes of the inertial frames of reference
in such way that if in one such frame of reference the beam of light has the 
direction $\vec{n}$, it will have the same direction in all inertial reference
frames \cite{70}. The resulting theory will be not the very special relativity
but the ordinary special relativity under the indicated special convention 
about the spatial axes orientations. This is true if we can choose $n^\mu$ 
arbitrarily and all such choices are equivalent. In fact, it is sufficient to 
require that the choices $\vec{n}$ and  $-\vec{n}$ are equivalent. Indeed, 
when  $\vec{n}\to -\vec{n}$, then $\vec{m}\to -\vec{m}$ in (\ref{eq102}) and, 
since $R^\mu_{\;\lambda}(-\vec{m},\alpha)R^\lambda_{\;\nu}(\vec{m},\alpha)=
\delta^\mu_\nu$, if we require the symmetry under rotations about $-\vec{m}$
and if $b=0$, we can easily transform the generalized Lorentz transformations 
(\ref{eq115}) into the ordinary Lorentz transformations by an additional 
rotation $x^{\prime\prime\,\mu}=R^\mu_{\;\nu}(-\vec{m},\alpha)x^{\prime\,\nu}$.
 
Therefore, if the space-time is Minkowskian, the very special relativity 
(ISIM(2) symmetry) is not natural in light of relativity principle and will
indicate its breaking, although very subtle, if it is indeed realized in 
nature. In our opinion far more natural possibility which completely 
respects the relativity principle, is that the space-time is Finslerian with 
the Lalan-Alway-Bogoslovsky metric (\ref{eq92}) but the parameter $b$ of this 
metric is very small. This conclusion is reinforced by the following analogy 
with the cosmological constant.

A surprising fact about Minkowski's ``Raum und Zeit'' lecture is that it
never mentions Klein's Erlangen program of defining a geometry by its
symmetry group \cite{18}. A link between Minkowski's presentation of special
relativity and Erlangen program was immediately recognized by Felix Klein
himself \cite{93} who remarked: ``What the modern physicists call the theory 
of relativity is the theory of invariants of the 4-dimensional space-time 
region $x,y,z,t$ (the Minkowski 'world') under a certain group of 
collineations, namely, the 'Lorentz group' ''. Untimely death of Minkowski
presumably hindered the appreciation of this important fact by physicists.
Only in 1954 Fantappi\'{e} rediscovered the connection and put forward an idea
which he himself calls ``an Erlangen program for physics'': a classification 
of possible physical theories through their group of symmetries \cite{94,95}. 

Interestingly, we have one more would-be ``missed opportunities'' story here 
\cite{I-3}. Fantappi\'{e} discovered that the ``final relativity'' group is 
not the Poincar\'{e} group but the De Sitter group, the Poincar\'{e} group 
being just a limit of the latter when the radius of curvature of the 
De Sitter space-time goes to infinity, much like the Galilei group being 
a limit of the Lorentz group when the speed of light goes to infinity. In fact,
the full story of possible kinematic groups and their interconnections under 
different limits was given later by H.~Bacry and J.~L\'{e}vy-Leblond 
\cite{96}. This picture seems quite natural in light of the Erlangen program 
and of ideas of group contractions and deformations put forward by  
In\"{o}n\"{u} and Wigner \cite{89} and Irving Segal \cite{97} about the same 
time Fantappi\'{e} found his ``final relativity''. Dyson argues \cite{I-3} 
that Minkowski, or any other pure mathematician, were in a position to 
conjecture already in 1908 that the true invariance group of the universe 
should be the de Sitter group rather than the Poincar\'{e} group and thus 
suggest a cosmological space expansion much earlier than Hubble discovered it, 
if only they would have brought Minkowski's argument to its logical conclusion.
 However, ``to be honest, it would be rather anachronistic to expect an 
approach like the one of Fantappi\'{e} at the beginning of the 20th century, 
as Dyson advocates'' \cite{95}. Even Fantappi\'{e} himself ``missed the chance 
to anticipate the theory of Lie group contractions and deformations'' 
\cite{95}.

Segal's principle  \cite{97,98,99} states that a true physical theory should 
be stable against small deformations of its underlying algebraic (group) 
structure (a nice informal review of Lie algebra contractions and deformations
can be found in \cite{100}). Lie algebra of inhomogeneous Galilei group is not
stable and its deformation leads to Lie algebra of  the Poincar\'{e} group.
Consequently, the relativity theory based on the Poincar\'{e} group has a 
greater range of validity than Galilean relativity. However, the Poincar\'{e}
Lie algebra is by itself unstable and its deformation leads to either de Sitter
or anti-de Sitter Lie algebras \cite{96}. Therefore, it is not surprising
that the cosmological constant turned out to be not zero and correspondingly,
the asymptotic vacuum space-time is not Minkowski but de Sitter space-time.
What is really surprising is why the cosmological constant is so small that
makes special relativity valid for all practical purposes.

Lie algebra of the very special relativity symmetry group ISIM(2) is not 
stable against small deformations of its structure and a physically relevant 
deformation DISIM$_b$(2) of it does exist \cite{88}. Therefore, in light of
Segal's principle, we expect that the very special relativity cannot be a true
symmetry of nature and should be replaced by DISIM$_b$(2) and the 
corresponding Finslerian space-time.  Drawing an analogy with the cosmological 
constant, it can be argued that $b$ is really not zero but very small. In 
this case, to detect the effects of Finslerian nature of  space-time in 
laboratory experiments will be almost impossible \footnote{In \cite{100A} 
a different Finslerian modification of space-time and its effect on 
the Maxwell's equations were considered along with the corresponding 
experimental consequences due to modifications of the Coulomb potential.
Although we expect that the true modification of the Minkowski space-time
is the Lalan-Alway-Bogoslovsky metric with the parameter $b$ so small that
probably it will be impossible to detect its presence in laboratory 
experiments, the ultimate judge is real experiments of course and any 
experimental efforts to place bounds on proposed Finsler parameters are 
worthwhile.}. Nevertheless, the question of the true value of the parameter 
$b$ has the same fundamental significance as the question why the cosmological 
constant is so small \cite{88}. Perhaps, both questions are just different 
parts of the same mystery.

\section{Concluding remarks}
In parallel to the advance in modern physics, in the middle of the twentieth 
century it became increasingly evident that Poincar\'{e}'s contribution to 
relativity was unjustly downplayed. As a result, some attempts to restore the 
justice followed. Unfortunately it was forgotten in the majority of these 
attempts that ``injustice cannot be corrected by committing another 
injustice'' \cite{101} and recurrent attempts to deny Einstein the discovery 
of special relativity took place \cite{102}. A great amount of these attempts
don't deserve even to be mentioned. However, there are some of them that are 
worth our attention and which are desirable to be clarified.

Renowned English mathematician Edmund Whittaker wrote a book {\em A History 
of the Theories of Aether and Electricity}, first edition of which was 
published in 1910. This is an  excellent and very detailed book describing 
the development of field theories from Descartes up to the year 1900. In the
final chapter of the book there is some survey of the work done after 1900.
It includes a note about Minkowski's {\em Raum und Zeit} paper. Then little
known Einstein is mentioned twice in footnotes \cite{103}. First footnote
says that Einstein completed Lorentz's work on Lorentz transformations. Then
the footnote cites Voigt and says that ``It should be added that the 
transformation in question had been applied to the equation of vibratory 
motions many years before by Voigt'' \cite{103} that is not quite correct as 
we have explained earlier. The next footnote correctly attributes to 
Einstein the first clear expression of the opinion that all inertial frames 
are equivalent and no one is granted a primacy by having an absolute relation 
to the {\ae}ther. Poincar\'{e} is mentioned three times, also in footnotes, 
but never in the context of relativity.

Forty-three years later, in 1953, Whittaker published a revised edition of his 
book, which now included  a second volume covering the years from 1900 to 
1926 \cite{103A}. The chapter devoted to relativity was called ``The 
relativity theory of Poincar\'{e} and Lorentz'' and its content left little
doubt that Whittaker attributes the whole credit for the discovery of special 
relativity  almost exclusively to Poincar\'{e} and Lorentz. About Einstein's 
groundbreaking 1905 paper it was said that ``Einstein published a paper which 
set forth the relativity theory of Poincare and Lorentz with some 
amplifications, and which attracted much attention''. Why such injustice to
Einstein? Maybe the following long excerpt from Max Born's 26 September, 1953
letter to Einstein can shed some light on this difficult question.

``Very often I feel the need to write to you, but I usually suppress it to 
spare you the trouble of replying. Today, though, I have a definite reason 
--- that Whittaker, the old mathematician, who lives here as Professor 
Emeritus and is a good friend of mine, has written a new edition of his old 
book History of the Theory of the Ether, of which the second volume has 
already been published. Among other things it contains a history of the theory 
of relativity which is peculiar in that Lorentz and Poincar\'{e} are credited 
with its discovery while your papers are treated as less important. Although 
the book originated in Edinburgh, I am not really afraid you will think that 
I could be behind it. As a matter of fact I have done everything I could 
during the last three years to dissuade Whittaker from carrying out his plan, 
which he had already cherished for a long time and loved to talk about. 
I re-read the originals of some of the old papers, particularly some rather 
off-beat ones by Poincar\'{e}, and have given Whittaker translations of German 
papers (for example, I translated many pages of Pauli's Encyclopedia article 
into English with the help of my lecturer, Dr. Schlapp, in order to make it 
easier for Whittaker to form an opinion). But all in vain. He insisted that 
everything of importance had already been said by Poincar\'{e}, and that 
Lorentz quite plainly had the physical interpretation. As it happens, I know 
quite well how skeptical Lorentz was and how long it took him to become 
a relativist. I have told Whittaker all this, but without success'' 
\cite{104}. 

Born's role in this affair is more ambivalent that it can appear from this 
letter \cite{105}. At first sight Born was in a good position to be 
a competent judge in the relativity priority dispute. He recollects \cite{106}
that at the time when Einstein's  1905 paper appeared he was ``in Gottingen 
and well acquainted with the difficulties and puzzles encountered in the study 
of electromagnetic and optical phenomena in moving bodies, which we thoroughly 
discussed in a seminar held by Hilbert and Minkowski. We studied the recent 
papers by Lorentz and Poincar\'{e}, we discussed the contraction hypothesis 
brought forward by Lorentz and Fitzgerald, and we knew the transformations now 
known under Lorentz's name. Minkowski was already working on his 
4-dimensional representation of space and time, published in 1907, which 
became later the standard method in fundamental physics'' and all this was 
before Born became aware of Einstein's 1905 paper. Therefore, Born was well 
versed in pre-Einstein relativity of Lorentz and Poincar\'{e}. In spite of 
this, Einstein's paper greatly impressed Born. In his book {\em Physics in my 
generation} \cite{106} he wrote: 
 
``A long time before I read Einstein's famous 1905 paper, I knew the formal 
mathematical side of the special theory of relativity through my teacher 
Hermann Minkowski. Even so, Einstein's paper was a revelation to me which had 
a stronger influence on my thinking than any other scientific experience\ldots 
Einstein's simple consideration, by which he disclosed the epistemological 
root of the problem\ldots made an enormous impression, and I think it right 
that the principle of relativity is connected with his name, though Lorentz 
and Poincar\'{e} should not be forgotten'' \cite{105,106}.

Born's letter to Einstein leaves an impression that ``Born's attention was 
drawn to Poincar\'{e}'s papers by Whittaker, rather than the other way around, 
in contradiction to Born's recollection of having read Poincar\'{e}'s papers 
in his student days before reading Einstein's paper'' \cite{105}. We think the
contradiction is only apparent here. There is little doubt that Poincar\'{e}'s
papers were discussed at Minkowski's seminars and they were well known to
Born. The only explanation why he needed to re-read them in fifties to discover
that several of Poincar\'{e}'s pre-1905 statements sound very much like 
special relativity is that in 1905 or before Poincar\'{e}'s papers have not 
impressed him as much as Einstein's 1905 paper which was ``a revelation'' to 
him. Whittaker also was not aware of the importance of Poincar\'{e}'s papers 
in 1910. Only in fifties he was in a position to acknowledge their importance.

It is not surprising that modern physicists like, for example, Logunov 
\cite{9}, find Poincar\'{e}'s papers important and rise doubts about 
Poincar\'{e}'s priority in discovery of special relativity. Even Born, 
somewhat contrary to what he writes in the letter to Einstein, hesitated in 
giving the priority: ``Does this mean that Poincar\'{e} knew all this before 
Einstein? It is possible, but the strange thing is that this lecture 
[Poincar\'{e}'s 1909 lecture {\em La m\'{e}canique nouvelle}]
definitely gives you the impression that he is recording Lorentz's work\ldots 
On the other hand, Lorentz himself never claimed to be the author of the 
principle of relativity'' \cite{105,106}. Further he writes about Einstein's 
1905 paper: ``The striking point is that it contains not a single reference to 
previous literature. It gives you the impression of quite a new venture. But 
that is, of course, as I have tried to explain, not true'' \cite{106}. And 
this is, of course, ``precisely the conclusion that Whittaker reached\ldots 
after long discussions of the subject with Born'' \cite{105}.

However, Whittaker's conclusion, implicitly shared by many modern phy\-sicists
who had taken a trouble to indeed read Poincar\'{e}'s papers, is wrong.  
This conclusion is based on the retrospective reading of Poincar\'{e}'s papers.
However, in retrospective reading of really deep papers, and there is no doubt
that Poincar\'{e}'s relativity papers are really very deep, the reader can see
much more than it was possible to see by contemporaries, including the author,
who were bound by concrete historical context and prejudices of those days.
If we inspect the physical literature of the founding period of
relativity (1905-1918), we will clearly see that the scientific community 
never hesitated to give Einstein a due credit, and there was no such a thing 
as Poincar\'{e}'s version of relativity ``accessible and perceived as such by 
physicists of the times'' \cite{26}. It is a historical fact that the 
relativistic revolution, as seen in the contemporary physical literature
of those days, rightly or wrongly is dominated by only one name, the 
Einstein's papers playing the major role, while Poincar\'{e}'s ones being 
left virtually unnoticed \cite{26}. The importance of these papers was 
recognized only later. In this respect, Poincar\'{e}-Einstein mystery
``is an artefact of projecting backward a particular reading of scientific 
papers that does not correspond to what the actors of the time saw in them''
\cite{26}. We cannot rewrite the history to restore justice towards 
Poincar\'{e}. This is not needed indeed. A scientific rehabilitation of
Poincar\'{e}'s contribution to relativity is an accomplished fact now. This
contribution, being far ahead of his time, is still alive and modern scholars
can still find inspiration in it. Isn't this a wonderful miracle? 

Interestingly, it seems like Born in his letter to Einstein is trying to 
provoke Einstein over the priority issue \cite{105}. He writes: ``I am annoyed
about this, for he is considered a great authority in the English speaking 
countries and many people are going to believe him'' \cite{104}. Einstein's 
response in his 12 October, 1953 letter was short and priceless in its wisdom:

``Don't lose any sleep over your friend's book. Everybody does what he 
considers right or, in deterministic terms, what he has to do. If he manages 
to convince others, that is their own affair. I myself have certainly found 
satisfaction in my efforts, but I would not consider it sensible to defend 
the results of my work as being my own 'property', as some old miser might
defend the few coppers he had laboriously scraped together. I do not hold 
anything against him, nor of course against you. After all, I do not need to 
read the thing'' \cite{104}.

As witnessed by Robert Oppenheimer, Einstein in his last years `was 
a twentieth-century Ecclesiastes, saying with unrelenting and indomitable 
cheerfulness, 'Vanity of vanities, all is vanity' '' \cite{107}.
 
Of course we can wonder why Einstein, and not then much famous and educated 
Poincar\'{e}, was at the origin of relativistic revolution. Undoubtedly some 
peculiar personal character traits of Einstein played the role. ``He was 
almost wholly without sophistication and wholly without worldliness. I think 
that in England people would have said that he did not have much
'background' and in America that he lacked 'education'. This may throw some 
light on how these words are used. I think that this simplicity, this lack of 
clutter and this lack of cant, had a lot to do with his preservation 
throughout of a certain pure, rather Spinoza-like, philosophical monism, which 
of course is hard to maintain if you have been 'educated' and have a 
'background'. There was always with him a wonderful purity at once childlike 
and profoundly stubborn'' \cite{107}.

From the scientific point of view, the fundamental difference between the 
conceptual foundations of the Lorentz-Poincar\'{e} {\ae}ther based theory 
from one side, and Einstein's special relativity from another, played the 
crucial role. Most succinctly this difference was expressed by Lorentz 
himself: `the chief difference being that  Einstein simply 
postulates what we have deduced, with some difficulty, and not altogether 
satisfactorily, from the fundamental equations of the electromagnetic field''
\cite{108}. It is clear that in contrary to Einstein,  Lorentz and 
Poincar\'{e} considered special relativity as a derivative, not fundamental 
theory and tried to base it on a more general premises. Although very modern
in its spirit, such an approach had no chance in 1905. Even now we have little
clue how special relativity can be considered as an emergent and not a 
fundamental phenomenon.

Whittaker was not the only prominent mathematician who ill treated Einstein's
contribution to relativity. Outstanding Russian mathematician Vladi\-mir
Arnold also gives a quite distorted picture:

``Minkowski, being a teacher of  Einstein and a friend of Poincar\'{e}, had 
early suggested to Einstein that he should study Poincar\'{e}'s theory, and 
Einstein did (though never referring to this until a 1945 article)'' 
\cite{109}.

It is true that Minkowski was Einstein's former mathematics professor at the 
Federal Polytechnic School of Z\"{u}rich. However, at that time Einstein was 
less interested in abstract mathematics and often skipped Minkowski's classes 
during his studies at the Polytechnic \cite{21}. Needless to say, it is simply 
not true that Minkowski ever supervised Einstein in his scientific research. 
According to Arnold Sommerfeld's recollections ``Strangely enough no personal 
contacts resulted between his teacher of mathematics, Hermann Minkowski, and 
Einstein'' \cite{21}.

At first sight it seems difficult to explain why Poincar\'{e} and Einstein
almost did not refer to each other when writing about relativity. First we 
consider Poincar\'{e}'s silence which is, perhaps, easier to explain. As was
mentioned above, Poincar\'{e}'s objective was much more ambitious than
Einstein's as he wanted to derive special relativity as an emergent phenomenon.
It is quite possible therefore that Poincar\'{e} simply considered Einstein's
contribution as being too trivial in light of this bigger goal. 
``To Poincar\'{e}, Einstein's theory must have been seen as a poor attempt 
to explain a small part of the phenomena embraced by the Lorentz theory''
\cite{110}. Perhaps, later in his life Poincar\'{e} became aware of the
importance of  Einstein's work. In November of 1911, one year before his
unexpected death, when he was asked to write a recommendation letter for 
Einstein who was looking to become a professor of theoretical physics 
at Swiss Federal Institute of Technology in Zurich, Poincar\'{e} wrote:

``Monsieur Einstein is one of the most original minds I have known; in spite 
of his youth he already occupies a very honorable position among the leading 
scholars of his time. We must especially admire in him the ease with which he 
adapts himself to new concepts and his ability to infer all the consequences 
from them. He does not remain attached to the classical principles and, faced 
with a physics problem, promptly envisages all possibilities. This is 
translated immediately in his mind into an anticipation of new phenomena, 
susceptible some day to experimental verification. I would not say that all 
his expectations will resist experimental check when such checks will become 
possible. Since he is probing in all directions, one should anticipate, on the 
contrary, that most of the roads he is following will lead to dead ends; but, 
at the same time, one must hope that one of the directions he has indicated 
will be a good one; and that suffice'' \cite{20}. 

Einstein's silence is more difficult to explain. It is reasonable to assume
that young Einstein, like most other physicists of those times, just lacked
the proper mathematical education to dully appreciate Poincar\'{e}'s work.
``More surprisingly, Einstein continued to ignore Poincar\'{e}'s contribution 
in all his later writings on special relativity and in his autobiographical 
notes'' \cite{111}. For late Einstein it is impossible to assume that he
was not capable to recognize the importance of Poincar\'{e}'s work. How can we
then explain Einstein's silence? We think he simply had not read 
Poincar\'{e}'s key papers on relativity until perhaps later times.
Contrary to what can be expected, most scholars do not have a habit to read
the papers  of their fellow scientists attentively. Empirical studies of 
misprint distributions in citations indicate that only about 20\% of 
scientists read the original that they cite \cite{112}. In case of Einstein
we have his own confession to Ehrenfest: ``My dear friend, do you think that 
I make a habit of reading papers written by others?'' \cite{113}. In the
early 1950s Abraham Pais asked Einstein how Poincar\'{e}'s Palermo paper had 
affected his thinking. Einstein answered that he had never read the 
paper. Then Pais lent to Einstein his own copy of the paper which afterwords
vanished \cite{20}. ``Perhaps he did read it. In 1953 Einstein received an 
invitation to attend the forthcoming Bern celebration of the fiftieth 
anniversary of special relativity. Einstein wrote back that his health did 
not permit him to plan such a trip. In this letter Einstein mentions for 
the first time (as far as I know) Poincar\'{e}'s role in regard to the special 
theory: 'I hope that one will also take care on that occasion to honor 
suitably the merits of H.~A.~Lorentz and H.~Poincar\'{e}' '' \cite{20}.
Besides, two months before his death, Einstein wrote to his biographer Carl 
Seelig that Poincar\'{e} further deepened Lorentz's insight about the 
essential role of Lorentz's transformations for the analysis of Maxwell's 
equations \cite{111}.

There is still another aspect which makes Einstein-Poincar\'{e} priority
dispute pointless. Modern understanding of relativity is significantly 
different from the one that was cultivated at the beginning of the twentieth 
century. Two examples are the notions of {\ae}ther and relativity of 
simultaneity which are often used in the priority dispute.

Einstein proponents stress that Poincar\'{e} never abandoned the {\ae}ther. 
Indeed, in 1912, a few months before his death, he gave a talk at
a conference  at the University of London in which he said: ``Everything 
happens as if time were a fourth dimension of space, and as if 
4-dimensional space resulting from the combination of ordinary space and 
of time could rotate not only around an axis of ordinary space in such a way 
that time were not altered, but around any axis whatever. For the comparison 
to be mathematically accurate, it would be necessary to assign purely 
imaginary values to this fourth coordinate of space'' \cite{114,115}. Taken
out of context, it is tempting to consider this statement as an evidence
that Poincar\'{e} was accepting the Einstein-Minkowski concept of space-time.
Especially when he continues: ``the essential thing is to notice that in the 
new conception space and time are no longer two entirely distinct entities 
which can be considered separately, but two parts of the same whole, two parts 
which are so closely knit that they cannot be easily separated'' 
\cite{114,115}. However, it is clear from the full text of his talk that  he 
is not at all ready to accept this new conception, as he further continues 
with the conclusion: ``What shall be our position in view of these new 
conceptions? Shall we be obliged to modify our conclusions? Certainly not; we 
had adopted a convention because it seemed convenient and we had said that 
nothing could constrain us to abandon it. Today some physicists want to adopt 
a new convention. It is not that they are constrained to do so; they consider 
this new convention more convenient; that is all. And those who are not of 
this opinion can legitimately retain the old one in order not to disturb their 
old habits. I believe, just between us, that this is what they shall do for 
a long time to come'' \cite{114,115}.

Therefore, clearly, ``Poincar\'{e} never believed in the physical relevance 
of the conceptual revolution brought by Einstein in the concept of time
(and extended by Minkowski to a revolutionary view of the physical meaning of
space-time)'' \cite{115}. In particular, for Poincar\'{e} the {\ae}ther, even
unobservable by physical experiments, remained an important conceptual element.

Usually this stubbornness of Poincar\'{e} with respect to the {\ae}ther is
considered as his weak point, as an evidence that he didn't really understand
relativity. It is historically true that the abolishment of the {\ae}ther
by Einstein played a crucial role and revolutionized physics. However, frankly 
speaking, in retrospect, when this revolution came to its logical end in
modern physics, we can equally well consider Poincar\'{e}'s attitude as
prophetical. 

As modern physics has progressed in the twentieth century, it became 
increasingly evident that the vacuum, the basic  state of quantum field 
theory, is anything but empty space. In fact, at present an {\ae}ther,
``renamed and thinly disguised, dominates the accepted laws of physics''
\cite{116}. It is clear that only ``intellectual inertia'' \cite{117}
prevents us from using historically venerable word ``{\ae}ther'' instead of
``vacuum state'' when referring to the states with such complex physical 
properties as vacuum states of modern quantum theories.

However, the {\ae}ther of modern physics is Lorentz invariant and hardly the 
one which Poincar\'{e} had in mind. Interestingly both in classical mechanical 
systems \cite{118} and in condensed matter physics \cite{119} an effective
Lorentz symmetry may arise from non-relativistic background physics. This is 
reminiscent of Lorentz-Poincar\'{e} program of deriving the Lorentz symmetry
rather than postulating it. This program was never completely abandoned. For 
most notable attempts in this direction see the books of Brown \cite{86} and 
J\'{a}nossy \cite{120}. Such attempts had, and still have, little chance of 
complete success. It is not excluded however that they will get a better 
chance when our knowledge of the Planck scale physics progresses. Therefore, 
in this respect Poincar\'{e}'s idea of the {\ae}ther in which an effective 
Lorentz symmetry emerges in some degrees of freedom at low energy limit is 
still alive and modern. 

Now about relativity of simultaneity. Relativity of simultaneity was considered
by Einstein himself and his contemporaries as the most crucial and 
revolutionary aspect of special relativity. For example Max Planck in his 
lectures given at Columbia University in 1909 (consequently published by him 
in 1910) so describes Einstein's ideas on space and time: ``It need scarcely 
be emphasized that this new conception of the idea of time makes the most 
serious demands upon the capacity of abstraction and the power of imagination 
of the physicist. It surpasses in boldness everything previously suggested in 
speculative natural phenomena and even in the philosophical theories of 
knowledge: non-euclidean geometry is child's play in comparison. And, moreover,
the principle of relativity, unlike non-euclidean geometry, which only comes 
seriously into consideration in pure mathematics, undoubtedly possesses a real
physical significance. The revolution introduced by this principle into the 
physical conceptions of the world is only to be compared in extent and depth 
with that brought about by the introduction of the Copernican System of the 
universe'' \cite{6A}.

As was already mentioned above, relativity of simultaneity was originally
Poinca\-r\'{e}'s invention. Surprisingly, he did not attach a particularly
fundamental importance to it \cite{6A}, as witnessed by Poincar\'{e}'s 
conclusion in the chapter {\it The measure of time} of his book {\it The Value 
of Science} first published in 1905. We provide here this conclusion in its
entirety because of its importance in revealing Poincar\'{e}'s position
(the fragment is also reproduced in \cite{6A}):

``To conclude: we do not have a direct intuition of simultaneity, nor of the
the equality of two durations. If we think we have this intuition, this is 
an illusion. We replace it by the aid of certain rules which we apply almost 
always without taking count of them.

But what is the nature of these rules? No general rule, no rigorous rule; 
a multitude of little rules applicable to each particular case.

These rules are not imposed upon us and we might amuse ourselves in inventing 
others; but they could not be cast aside without greatly complicating the 
enunciation of the laws of physics, mechanics and astronomy.

We therefore choose these rules, not because they are true, but because they 
are the most convenient, and we may recapitulate them as follows: {\it The 
simultaneity of two events, or the order of their succession, the equality of 
two durations, are to be so defined that the enunciation of the natural laws 
may be as simple as possible. In other words, all these rules, all these 
definitions are only the fruit of an unconscious opportunism}'' \cite{120A}.

Poincar\'{e} proponents in the priority
dispute argue that Einstein synchronization, which Einstein himself considered
as the crucial element of special relativity, has in fact originated from 
Poincar\'{e}'s work. Above we had already commented on this difficult issue. 
Now we would like to consider another aspect of it. The fact is that in light 
of the modern view on relativity, Einstein synchronization can no longer be 
considered as a crucial element. What is really essential is Minkowski 
geometry (or slightly Finslerian) of space-time in the absence of gravity (and 
cosmological constant). Einstein synchronization is just a convention, or as  
Einstein himself put it, a ``stipulation'' which allows to introduce 
a convenient coordinate chart in Minkowski space-time. However, an 
introduction of any other coordinate charts, although probably less 
convenient, is equally possible. This modern view is very close to the above 
given Poincar\'{e}'s attitude.
 
For example, we can use 'everyday' clock 
synchronization when clocks are  adjusted by using time signal broadcasted by 
some radio-station telling that, for example, ``At the six stroke, the time 
will be 12 o'clock exactly'' \cite{121}. All coordinate-dependent quantities, 
including the Lorentz transformations too, can experience drastic changes 
under the change of synchronization convention. For example, such once thought 
intrinsic feature of relativity as relativity of simultaneity is no longer 
true when 'everyday' clock synchronization is used \cite{121}. Under such 
synchronization  the ``Lorentz'' transformations relating the two 'everyday' 
observers $S$ and $S^\prime$ with relative velocity $V$ (as measured in the 
frame $S$) along the $x$-axis have the following form \cite{121}
\begin{equation}
\begin{array}{l}
x^\prime=\frac{x-Vt}{\sqrt{1+2\beta}}, \\
y^\prime=y, \\
z^\prime=z, \\
t^\prime=\sqrt{1+2\beta}\,t.
\end{array}
\label{eq117}
\end{equation}
where $\beta=V/c$. It is clear from these transformations that simultaneity
under 'everyday' synchronization is absolute: if $\Delta t=0$ in some system
$S$ then in all other inertial systems $S^\prime$ we also shall have 
$\Delta t^\prime=0$.

Conventionality of simultaneity was systematically investigated by Reichenbach
(see, for example, \cite{122}). Suppose that two distant clocks $A$ and $B$ 
are motionless in a common inertial frame $S$. If a light signal is sent from 
$A$ at time $t_A$, is instantaneously reflected by $B$ at time $t_B$, 
and arrives back at $A$ at time $t_A^\prime$, then $A$ and $B$ are 
Poincar\'{e}-Einstein synchronized if $t_B-t_A=t_A^\prime-t_B$, or
\begin{equation}
t_B=\frac{1}{2}(t_A+t_A^\prime)=t_A+\frac{1}{2}(t_A^\prime-t_A).
\label{eq118}
\end{equation}
Reichenbach modifies this definition of synchronization as follows
\begin{equation}
t_B=t_A+\epsilon (t_A^\prime-t_A),
\label{eq119}
\end{equation}
where now $\epsilon$ is some real parameter and $0\le\epsilon\le 1$ (because
for causality reasons we need $t_B\ge t_A$ and $t_B\le t_A^\prime$), the 
equality corresponding to the infinite one-way velocity of the light signal.
In fact, these limiting cases were excluded by Reichenbach).

The modification of Lorentz transformations needed to meet this general
$\epsilon$-syn\-chro\-nization, was considered by Winnie \cite{123}. If 
$\epsilon$-synchronization is adopted in the reference frame $S$ while
$\epsilon^\prime$-synchronization is adopted in the reference frame 
$S^\prime$, Winnie transformations have the form \cite{123}
\begin{equation}
\begin{array}{l}
x^\prime=\frac{x-Vt}{\alpha}, \\
y^\prime=y, \\
z^\prime=z, \\
t^\prime=\frac{1}{\alpha}\left\{\left[1+2\beta(1-\epsilon-\epsilon^\prime)
\right]t-\left[2(\epsilon-\epsilon^\prime)+4\beta\epsilon(1-\epsilon)\right]
\frac{x}{c}\right\},
\end{array}
\label{eq120}
\end{equation}
where
\begin{equation}
\alpha=\sqrt{[1-(2\epsilon-1)\beta]^2-\beta^2}.
\label{eq121}
\end{equation}
If $\epsilon=\epsilon^\prime=1/2$, we recover the ordinary Lorentz 
transformations, and if $\epsilon=\epsilon^\prime=0$ we get the 
transformations (\ref{eq117}) which correspond to 'everyday' synchronization.

It is interesting to note that the 'everyday' synchronization is not the only 
choice that corresponds to absolute simultaneity. Historically Frank Robert 
Tangherlini was the first to realize that absolute simultaneity doesn't 
contradict special relativity in his 1958 PhD dissertation supervised by 
Sidney Drell and, at the initial stage of the work, Donald Yennie 
\cite{124,125,126}. Tangherlini's transformations have the form
\begin{equation}
\begin{array}{l}
x^\prime=\gamma(x-Vt), \\
y^\prime=y, \\
z^\prime=z, \\
t^\prime=\gamma^{-1}t,
\end{array}
\label{eq122}
\end{equation}
and they correspond to the special case of the Winnie transformations
(\ref{eq117}) when $\epsilon=1/2$ and $\epsilon^\prime=(1+\beta)/2$.
External synchronization provides a possible realization of the Tangherlini's
simultaneity \cite{42,126A}. First clocks in the ``{\ae}ther'' frame $S$ are 
Poincar\'{e}-Einstein synchronized with $\epsilon=1/2$. Then these clocks 
are used to synchronize nearby clocks in the moving frame $S^\prime$ by
adjusting clocks from $S^\prime$ to $t^\prime=0$ whenever they fly past 
a clock in $S$ which shows $t=0$.

Note that the Tangherlini transformations (\ref{eq122}) were obtained
twenty years before his thesis by English mathematician Albert Eagle,
although from erroneous anti-relativistic premises \cite{126A,126B} ---
one more example that correct mathematics doesn't guarantee a correct
physical interpretation.

Some other simultaneity conventions, not necessarily based on light signals 
and their two-way universal velocity $c$,  are also feasible. For examples,
inhabitants of the homogeneous isotropic ocean can find convenient to use 
for synchronization light signals with two-way velocity $c^\prime<c$ 
\cite{127}, and dolphins can even prefer sound waves for this goal 
\cite{128}. Of course, coordinate-independent quantities are not affected
by synchronization conventions. Various synchronization conventions are useful
to separate the effects due to synchronization and the real contraction of 
moving bodies and slowing down of moving clocks. Such investigations allow
us to come to the conclusion that ``the results of relativistic experiments 
have their origin in the length contraction and time dilatation effects which 
are so real as a change of the length of a rod caused by the change of 
temperature'' \cite{129,130}. 

Rosen's $c^\prime$-relativity gives an interesting perspective on superluminal 
objects \cite{127}. Effective $c^\prime$-relativity (as far as the 
electromagnetic phenomena are concerned) precludes charged particles in the 
Rosen's ocean to acquire speeds greater than or equal to $c^\prime$. However,
one might have an energetic cosmic muon enter the ocean from outside with
a speed greater than $c^\prime$. The behavior of such a particle could not
be described in the framework of $c^\prime$-relativity. For example, the
inhabitants of the ocean will observe  ``vacuum'' Cherenkov radiation emanated
from such a particle --- a phenomenon they certainly think impossible on
the base of $c^\prime$-relativity. It was speculated in \cite{131} that
likewise $c$-relativity might also not embrace some hidden sectors weakly 
coupled to our visible sector of the world. Then superluminal objects called 
elvisebrions in \cite{131} can enter the visible sector and their behavior 
could not be described in the framework of special relativity.

Tangherlini was guided by the logic of general relativity when undertaking
his investigation of absolute simultaneity \cite{126}. According to this 
logic, equations (\ref{eq122}) represent just a coordinate transformation 
which makes the metric tensor non-diagonal. Indeed, it is easy to get from 
(\ref{eq122}) that
\begin{equation}
dx=\gamma^{-1}dx^\prime+\gamma V\,dt^\prime,\;\;\;dt=\gamma \,dt^\prime,
\label{eq123}
\end{equation}
and the relativistic interval (line element) takes the form
\begin{equation}
ds^2=c^2\,dt^{\prime\,2}-2V\,dx^\prime dt^\prime -\gamma^{-2}dx^{\prime\,2}-
dy^{\prime\,2}-dz^{\prime\,2}.
\label{eq124}
\end{equation}
From the point of view of general relativity, these coordinates are as good
as any others because of general covariance. However, as was already shown by 
Kretschmann in 1917, any space-time theory can be expressed in
general covariant manner, not only general relativity. Therefore, the 
coordinates (\ref{eq122}) are as good as any others in special relativity
too. It is just an other expression of ``intellectual inertia'' that 
standard textbook presentations of special relativity are based solely on 
inertial observers and diagonal metric. Another misbelief, which still 
prevails, is that accelerated observers cannot be treated by special 
relativity, and that one needs general relativity for their incorporation. 
In fact, there is no principal difficulty to consider Minkowski space-time 
from the point of view of non-inertial observers at the expense of somewhat 
more sophisticated mathematics \cite{133} (for more traditional exposition of 
special relativity see, for example, an excellent textbook \cite{133A}).  

We hope that the above given examples show clearly how different the present 
day interpretation of key features of special relativity is from what had in 
mind founding fathers of this magnificent theory. And the story of special 
relativity is not finished yet. For example, the debate about conventionality 
of simultaneity still continues and seems to be far from being settled 
(see,  for instance, \cite{134} and references therein). The initial 
observer-centered presentation of special relativity was changed to the modern 
geometry-centered presentation. However, it seems that the full potential of 
the notion of an observer as ontologically prior to either space or space-time 
is not yet exhausted \cite{135,136} and the initial presentation of special 
relativity, although renewed and enriched by our experience with space-time 
picture, can still strike back.

In light of this immense and still continuing progress of modern physics,
attempts to retrospectively induce an artificial  Poincar\'{e}-Einstein
priority dispute and rewrite the history seem minute. We will be happy if 
this arid and futile dispute will come to its end. There is nothing 
scientific in it and its presence only emphasizes hideous traits of human 
nature. However, there is no single thing we can do, except writing this 
article, to help to end this unfortunate dispute. It remains, like aging 
Einstein, to repeat after  Ecclesiastes 'Vanity of vanities, all is vanity'.  

\section*{appendixes}
\appendix\normalsize
\section{Uniformly accelerated motion}
\label{AppA}
Uniformly accelerated motion is defined as 
having a zero jerk (time derivative of the acceleration 3-vector) in 
instantaneous rest frame \cite{51}. Differentiating velocity 4-vector
\begin{equation}
u^\mu=\gamma\,(c,\,\vec{V}),
\label{eq60}
\end{equation}
we get successively the acceleration 4-vector
\begin{eqnarray} &&
\dot u^\mu=\frac{du^\mu}{d\tau}=\gamma\,\frac{du^\mu}{dt}=
\gamma^4\,\left(\vec{\beta}\cdot\vec{a},\,
\frac{\vec{a}}{\gamma^2}+(\vec{\beta}\cdot\vec{a})\vec{\beta}\right )=
\nonumber \\ &&
\gamma^4\,\left(\vec{\beta}\cdot\vec{a},\,\vec{a}+\vec{\beta}\times
(\vec{\beta}\times\vec{a})\right ),
\label{eq61}
\end{eqnarray}
and the naive jerk 4-vector 
\begin{eqnarray} &&
\ddot u^\mu=\frac{d\dot u^\mu}{d\tau}=
\gamma^5\left(\frac{\vec{a}\cdot\vec{a}}{c}+\vec{\beta}\cdot\vec{b}+
\frac{4\gamma^2}{c}\,(\vec{\beta}\cdot\vec{a})^2,\;\frac{\vec{b}}{\gamma^2}+
3\,\frac{(\vec{\beta}\cdot\vec{a})}{c}\,\vec{a}+ \right .
\nonumber \\ && \left .
\left[\vec{\beta}\cdot\vec{b}+
\frac{\vec{a}\cdot\vec{a}}{c}+4\,\frac{\gamma^2}{c}\,
(\vec{\beta}\cdot\vec{a})^2\right]\vec{\beta}\right ),
\label{eq62}
\end{eqnarray}
where $\vec{a}=\frac{d\vec{V}}{dt}$ and $\vec{b}=\frac{d\vec{a}}{dt}$.
In the instantaneous rest frame
\begin{equation}
\ddot u^\mu_{(0)}=\left(\frac{\vec{a}_{(0)}\cdot\vec{a}_{(0)}}{c},\;
\vec{b}_{(0)}\right),
\label{eq63}
\end{equation}
which indicates that $\ddot u^\mu$ is not necessarily spacelike and could be 
nonzero even if 3-jerk satisfies $\vec{b}_{(0)}=0$. Due to these properties 
$\ddot u^\mu$ can not be really considered as a relativistic jerk 4-vector 
\cite{51A}. To find a better candidate for the role of the relativistic jerk 
4-vector let us note that
\begin{equation}
u_\mu\ddot u^\mu=\vec{a}_{(0)}\cdot\vec{a}_{(0)}=-\frac{\dot u_\nu\dot u^\nu}
{c^2}\,u_\mu u^\mu,
\label{eq64}
\end{equation}
which implies
\begin{equation}
u_\mu\left (\ddot u^\mu+\frac{\dot u_\nu\dot u^\nu}{c^2}\,u^\mu\right )=0.
\label{eq65}
\end{equation}
Therefore, the 4-vector
\begin{equation}
\Gamma^\mu=\ddot u^\mu+\frac{\dot u_\nu\dot u^\nu}{c^2}\,u^\mu
\label{eq66}
\end{equation}
is spacelike because it is orthogonal to the timelike 4-vector $u^\mu$.
Besides, in the instantaneous rest frame $\Gamma^\mu_{(0)}=(0,\;\vec{b}_{(0)}
)$. Hence, it makes more sense to define the relativistic jerk to be 
$\Gamma^\mu$ \cite{51A}. Then the covariant condition of uniformly accelerated 
motion is the vanishing of jerk 4-vector $\Gamma^\mu$ \cite{51} (in \cite{51} 
$\Gamma^\mu$ is called the Abraham 4-vector due to its relation to the 
Lorentz-Abraham radiation reaction force). Using
\begin{equation}
\dot u_\nu\dot u^\nu=-\gamma^4\left [\vec{a}\cdot\vec{a}+\gamma^2(\vec{\beta}
\cdot\vec{a})^2\right],
\label{eq67}
\end{equation}
which follows from (\ref{eq61}), and
\begin{equation}
\vec{b}+3\,\frac{\gamma^2}{c}\,(\vec{\beta}\cdot\vec{a})\,\vec{a}=
\frac{1}{\gamma^3}\,\frac{d}{dt}(\gamma^3\,\vec{a}),
\label{eq68}
\end{equation}
we get after some algebra
\begin{eqnarray} &&
\Gamma^\mu=\gamma^2\left(\vec{\beta}\cdot\frac{d}{dt}(\gamma^3\,\vec{a}),\;
\frac{1}{\gamma^2}\,\frac{d}{dt}(\gamma^3\,\vec{a})+\left (\vec{\beta}\cdot
\frac{d}{dt}(\gamma^3\,\vec{a})\right )\,\vec{\beta}\right )=
\nonumber \\ &&
\gamma^2\left(\vec{\beta}\cdot\frac{d}{dt}(\gamma^3\,\vec{a}),\;
\frac{d}{dt}(\gamma^3\,\vec{a})+\vec{\beta}\times\left(\vec{\beta}\times
\frac{d}{dt}(\gamma^3\,\vec{a})\right )\right  ).
\label{eq69}
\end{eqnarray}
Therefore, a uniformly accelerated motion is characterized by the condition
\cite{51}
\begin{equation}
\frac{1}{\gamma^3}\,\frac{d}{dt}(\gamma^3\,\vec{a})=
\vec{b}+3\,\frac{\gamma^2}{c}\,(\vec{\beta}\cdot\vec{a})\,\vec{a}=0,
\label{eq70}
\end{equation}
which is equivalent to: 
\begin{equation}
\gamma^3\,\vec{a}=\vec{g},
\label{eq71}
\end{equation}
with some constant vector $\vec{g}$.

Suppose we want to interpret special conformal transformation (\ref{eq58})
as a transition to a new reference frame. How does the origin 
$\vec{r}^{\,^\prime}=0$ of this frame move? When $\vec{r}^{\,^\prime}=0$,
we will have from (\ref{eq58})
\begin{equation}
\vec{r}=\vec{\alpha}\,x\cdot x,
\;\;\;x\cdot x=x_\mu\,x^\mu=c^2t^2-\vec{r}\cdot\vec{r}.
\label{eq72}
\end{equation}
Differentiating this relation we find 
\begin{equation}
\vec{V}=2\vec{\alpha}\,(c^2t-\vec{r}\cdot\vec{V}),\;\;
\vec{a}=2\vec{\alpha}\,(c^2-V^2-\vec{r}\cdot\vec{a})=\frac{2\vec{\alpha}}
{\gamma^2}(c^2-\gamma^2\,\vec{r}\cdot\vec{a}).
\label{eq73}
\end{equation}
On the other hand, taking the scalar products of (\ref{eq73}) and (\ref{eq72}),
we get the relations
\begin{equation}
\vec{r}\cdot\vec{V}=2\,\vec{\alpha}\cdot\vec{\alpha}\,x\cdot x\,
(c^2t-\vec{r}\cdot\vec{V}),\;\;\vec{r}\cdot\vec{a}=2\,\vec{\alpha}\cdot
\vec{\alpha}\,x\cdot x\left(\frac{c^2}{\gamma^2}-\vec{r}\cdot\vec{a}\right ).
\label{eq74}
\end{equation}
From these equations $\vec{r}\cdot\vec{V}$ and $\vec{r}\cdot\vec{a}$ can be
determined, and substituting the results back in (\ref{eq73}) we get
\begin{equation}
\vec{V}=\frac{2\vec{\alpha}c^2t}{1+2\,\vec{\alpha}\cdot\vec{\alpha}\;
x\cdot x},\;\;\;
\vec{a}=\frac{2\vec{\alpha}c^2}{\gamma^2}\,\frac{1}{1+2\,\vec{\alpha}\cdot
\vec{\alpha}\;x\cdot x}.
\label{eq75}
\end{equation}
Therefore,
\begin{equation}
1-\beta^2=1-\frac{4\,\vec{\alpha}\cdot\vec{\alpha}\,c^2t^2}{(1+2\,\vec{\alpha}
\cdot\vec{\alpha}\;x\cdot x)^2}=\frac{1}{(1+2\,\vec{\alpha}\cdot\vec{\alpha}
\;x\cdot x)^2},
\label{eq76}
\end{equation}
where we have taken into account that in light of (\ref{eq72})
\begin{equation}
c^2t^2=x\cdot x+\vec{r}\cdot\vec{r}=x\cdot x+\vec{\alpha}\cdot\vec{\alpha}\;
(x\cdot x)^2.
\label{eq77}
\end{equation}
As we see,
\begin{equation}
\gamma=1+2\,\vec{\alpha}\cdot\vec{\alpha}\;x\cdot x,
\label{eq78}
\end{equation}
and then the second equation in (\ref{eq75}) gives 
$\gamma^3\,\vec{a}=2c^2\vec{\alpha}$ what implies that the point 
$\vec{r}^{\,\prime}=0$ experiences uniformly accelerated motion.

\section{Lalan-Alway-Bogoslovsky transformations}
\label{AppB}
Let us consider an inertial reference frame $S^\prime$ 
which moves with the velocity $\vec{V}$. If for a moment we consider 
$n^\mu$ as a normal 4-vector, not the fixed one, we may write
\begin{equation}
\left(n_\nu dx^\nu\right)^{2b}\left(dx_\mu dx^\mu\right)^{1-b}=
\left(n^\prime_\nu dx^{\prime\nu}\right)^{2b}\left(dx^\prime_\mu 
dx^{\prime\mu}\right)^{1-b},
\label{eq93}
\end{equation} 
where
\begin{equation}
n_0^\prime=\gamma\left(n_0-\beta\,\frac{\vec{n}\cdot\vec{V}}{V}\right)=
\gamma(1-\vec{n}\cdot\vec{\beta}),\;\;\vec{\beta}=\frac{\vec{V}}{c},
\label{eq94}
\end{equation} 
and
\begin{eqnarray} &&
\vec{n}^\prime=\frac{\vec{V}}{V}\,\gamma\left(\frac{\vec{n}\cdot\vec{V}}{V}-
\beta n_0\right)+\left(\vec{n}-\frac{\vec{n}\cdot\vec{V}}{V}\,\frac{\vec{V}}
{V}\right)= \nonumber \\ &&
\vec{n}+\frac{\vec{V}}{V}\left[(\gamma-1)\frac{\vec{n}\cdot\vec{V}}
{V}-\gamma\beta\right].
\label{eq95}
\end{eqnarray}
Using
\begin{equation}
\gamma-1=\frac{\gamma^2-1}{\gamma+1}=\frac{\gamma^2\beta^2}{\gamma+1},
\label{eq96}
\end{equation}
we can rewrite (\ref{eq95}) as follows
\begin{equation}
\vec{n}^\prime=\vec{n}-\gamma\vec{\beta}\left[1-\frac{\gamma}{\gamma+1}\,
\vec{n}\cdot\vec{\beta}\right]=\vec{n}-\frac{\gamma\vec{\beta}}{\gamma+1}
\left[1+\gamma(1-\vec{n}\cdot\vec{\beta})\right].
\label{eq97}
\end{equation}
If $\vec{V}$ is not parallel to $\vec{n}$, $\vec{n}^\prime$ will be not
parallel to $\vec{n}$ neither. If $\alpha$ is an angle between $\vec{n}
^\prime$ and $\vec{n}$, we will have
\begin{equation}
\sin{\alpha}=\frac{|\vec{n}\times\vec{n}^\prime|}{n_0^\prime}=
\frac{|\vec{n}\times\vec{\beta}|[1+\gamma(1-\vec{n}\cdot\vec{\beta})]}
{(\gamma+1)(1-\vec{n}\cdot\vec{\beta})},
\label{eq98}
\end{equation}
where we have used that $|\vec{n}^\prime|=n_0^\prime$ because the Lorentz
transformation leaves $n_\mu^\prime$ lightlike. Alternatively, we can write
\begin{eqnarray} &&
\cos{\alpha}=\frac{\vec{n}\cdot\vec{n}^\prime}{n_0^\prime}=
\frac{1-\gamma(\vec{n}\cdot\vec{\beta}-1+1)+\frac{\gamma^2}{\gamma+1}\,
(\vec{n}\cdot\vec{\beta})^2}{\gamma(1-\vec{n}\cdot\vec{\beta})}=
\nonumber \\ &&
1+\frac{1-\gamma+\frac{1}{\beta^2}\,(\gamma-1)(\vec{n}\cdot\vec{\beta})^2}
{\gamma(1-\vec{n}\cdot\vec{\beta})}=1-\frac{\gamma-1}{\gamma\beta^2}\,
\frac{\beta^2-(\vec{n}\cdot\vec{\beta})^2}{1-\vec{n}\cdot\vec{\beta}}.
\label{eq99}
\end{eqnarray}
But $\beta^2-(\vec{n}\cdot\vec{\beta})^2=[\vec{n}\times\vec{\beta}]^2$, and
finally
\begin{equation}
\cos{\alpha}=1-\frac{\gamma-1}{\gamma\beta^2}\,\frac{[\vec{n}\times
\vec{\beta}]^2}{1-\vec{n}\cdot\vec{\beta}}=1-\frac{\gamma}{\gamma+1}\,
\frac{[\vec{n}\times\vec{\beta}]^2}{1-\vec{n}\cdot\vec{\beta}}.
\label{eq100}
\end{equation}
We can make $\vec{n}^\prime$ again parallel to $\vec{n}$ by an additional
rotation around the axis $\vec{n}^\prime\times\vec{n}\parallel \vec{n}\times
\vec{\beta}$ by certain angle $\alpha$. For the radius vector $\vec{r}$ the 
result of this additional rotation is given by the Euler-Rodrigues formula 
\cite{67} (for proofs of this formula see, for example, \cite{68,69}) 
\begin{equation}
\vec{r}^{\,\prime\prime}=\vec{r}^{\,\prime}+(\vec{m}\times\vec{r}^{\,\prime})
\sin{\alpha}+[\vec{m}\times(\vec{m}\times\vec{r}^{\,\prime})](1-\cos{\alpha}),
\label{eq101}
\end{equation}
where
\begin{equation}
\vec{m}=\frac{\vec{n}\times\vec{\beta}}{|\vec{n}\times\vec{\beta}|}
\label{eq102}
\end{equation}
is the unit vector along the axis of the rotation and
\begin{equation}
\vec{r}^{\,\prime}=\vec{r}+\frac{\vec{\beta}}{\beta}\left[(\gamma-1)\,\frac{
\vec{r}\cdot\vec{\beta}}{\beta}-\gamma\beta\,x_0\right]
\label{eq103}
\end{equation}
is the result of the preceding Lorentz transformation on the spatial part
of the $x^\mu$ 4-vector. Using (\ref{eq98}) and (\ref{eq100}) along with
\begin{equation}
\vec{m}\times\vec{r}^{\,\prime}=\frac{(\vec{n}\times\vec{\beta})\times
\vec{r}^{\,\prime}}{|\vec{n}\times\vec{\beta}|}=\frac{\vec{\beta}(\vec{n}
\cdot\vec{r}^{\,\prime})-\vec{n}(\vec{\beta}\cdot\vec{r}^{\,\prime})}
{|\vec{n}\times\vec{\beta}|}
\label{eq104}
\end{equation}
and
\begin{eqnarray} &&
\vec{m}\times(\vec{m}\times\vec{r}^{\,\prime})=\frac{(\vec{n}\times
\vec{\beta})\times[\vec{\beta}(\vec{n}\cdot\vec{r}^{\,\prime})-\vec{n}
(\vec{\beta}\cdot\vec{r}^{\,\prime})]}{[\vec{n}\times\vec{\beta}]^2}=
\nonumber \\ &&
\frac{\vec{\beta}\,[(\vec{n}\cdot\vec{r}^{\,\prime})(\vec{n}\cdot\vec{\beta})-
\vec{\beta}\cdot\vec{r}^{\,\prime}]+\vec{n}\,[(\vec{\beta}\cdot\vec{r}^{\,
\prime})(\vec{\beta}\cdot\vec{n})-\beta^2(\vec{n}\cdot\vec{r}^{\,\prime})]}
{[\vec{n}\times\vec{\beta}]^2},
\label{eq105}
\end{eqnarray}
we get
\begin{eqnarray} &&
\vec{r}^{\,\prime\prime}=\vec{r}^{\,\prime}+\frac{\vec{\beta}}{1-\vec{n}\cdot
\vec{\beta}}\left[\vec{n}\cdot\vec{r}^{\,\prime}-\frac{\gamma}{\gamma+1}\,
\vec{\beta}\cdot\vec{r}^{\,\prime}\right]+
\nonumber \\ &&
\frac{\vec{n}}{1-\vec{n}\cdot
\vec{\beta}}\left[(\vec{\beta}\cdot\vec{r}^{\,\prime})\left(\frac{2\gamma}
{\gamma+1}\,\vec{\beta}\cdot\vec{n}-1\right)-
\frac{\gamma-1}{\gamma}\,
\vec{n}\cdot\vec{r}^{\,\prime}\right].
\label{eq106}
\end{eqnarray}
This result coincides with the one that follows from Eq.34 of \cite{70}. As an
other check, if we take $\vec{r}^{\,\prime}=\vec{n}^\prime$ and use the
relations
\begin{equation}
\vec{\beta}\cdot\vec{n}^\prime=\gamma[\vec{\beta}\cdot\vec{n}-\beta^2],\;\;\;
\vec{n}\cdot\vec{n}^\prime=1-\gamma\,\vec{\beta}\cdot\vec{n}+\frac{\gamma^2}
{\gamma+1}\,(\vec{\beta}\cdot\vec{n})^2
\label{eq107}
\end{equation}
that follow from (\ref{eq97}), we get after some calculations the correct 
result
\begin{equation}
\vec{n}^{\prime\prime}=\gamma(1-\vec{\beta}\cdot\vec{n})\,\vec{n}.
\label{eq108}
\end{equation}
Consequently, after the combined transformations
\begin{equation}
x^{\prime\prime\mu}=R^\mu_{\;\nu}(\vec{m};\alpha)\,L^\nu_{\;\sigma}(\vec{V})
x^\sigma,\;\;\;n^{\prime\prime\mu}=R^\mu_{\;\nu}(\vec{m};\alpha)\,L^\nu_{
\;\sigma}(\vec{V})n^\sigma,
\label{eq109}
\end{equation}
where the matrix $R^\mu_{\;\nu}(\vec{m};\alpha)$ represents a rotation by the
angle $\alpha$ around the vector $\vec{m}$ and the matrix $L^\nu_{\;\sigma}
(\vec{V})$ corresponds to a Lorentz boost with the velocity $\vec{V}$,
we will obtain
\begin{eqnarray} &&
\left(n_\nu dx^\nu\right)^{2b}\left(dx_\mu dx^\mu\right)^{1-b}=
\left(n^{\prime\prime}_\nu dx^{\prime\prime\nu}\right)^{2b}
\left(dx^{\prime\prime}_\mu dx^{\prime\prime\mu}\right)^{1-b}=
\nonumber \\ &&
\left[\gamma(1-\vec{\beta}\cdot\vec{n})\right]^{2b}\left(n_\nu 
dx^{\prime\prime\nu}\right)^{2b}\left(dx^{\prime\prime}_\mu 
dx^{\prime\prime\mu}\right)^{1-b}.
\label{eq110}
\end{eqnarray} 
The unwanted conformal factor $\left[\gamma(1-\vec{\beta}\cdot\vec{n})
\right]^{2b}$ can be compensated by final dilatation
\begin{equation}
x^{\prime\prime\prime\mu}=D(\lambda)x^{\prime\prime\mu}\equiv
\lambda\,x^{\prime\prime\mu},
\label{eq111}
\end{equation} 
with the suitably chosen scale-factor 
\begin{equation}
\lambda=\left[\gamma(1-\vec{\beta}\cdot\vec{n})\right]^b, 
\label{eq112}
\end{equation}
and we finally get the desired invariance
\begin{equation}
\left(n_\nu dx^\nu\right)^{2b}\left(dx_\mu dx^\mu\right)^{1-b}=
\left(n_\nu dx^{\prime\prime\prime\nu}\right)^{2b}\left(
dx^{\prime\prime\prime}_\mu dx^{\prime\prime\prime\mu}\right)^{1-b}.
\label{eq113}
\end{equation}

\section*{Acknowledgements}
An occasional remarks of A.I. Milstein and V.~I.~Telnov about  Vladi\-mir
Arnold's account of Poincar'{e}'s role in the development of special 
relativity triggered our interest and led to this investigation. Some 
constructive comments from our collegues helped us to make the manuscript 
more clear. The work of Z.K.S. is supported by the Ministry of Education 
and Science of the Russian Federation.

\end{document}